\documentclass[aps,twocolumn,preprintnumbers,amsmath,amssymb,nofootinbib,superscriptaddress,notitlepage]{revtex4}
\usepackage{graphicx,color,dcolumn,booktabs,bm}
\usepackage{longtable,lscape}
\usepackage{txfonts}
\usepackage{overpic}
\usepackage{amssymb}
\usepackage{gensymb}
\usepackage{indentfirst}
\usepackage{feynmf}
\usepackage{slashed}
\usepackage{cases}
\usepackage{multirow}
\usepackage{appendix}
\usepackage{float}
\usepackage[figuresright]{rotating}
\usepackage{subfigure}
\usepackage[colorlinks,
citecolor=blue,
anchorcolor=red,
menucolor=red,
linkcolor=red,
filecolor=red,
runcolor=red,
urlcolor=blue,
frenchlinks=red]{hyperref}
\usepackage{epstopdf}
\usepackage{bm}
\usepackage{tabularx}
\usepackage{threeparttable}
\usepackage{bbding}
\usepackage{array}
\usepackage[section]{placeins}
\usepackage{makecell}
\usepackage{comment}
\makeatletter

\newcommand{\Rmnum}[1]{\expandafter\@slowromancap\romannumeral #1@}
\makeatother

\usepackage{soul}

\begin{document}

\title{Decoding the role of $\rho$ mesonic states for elucidating the $e^+e^-\to a_2(1320)\pi$ data and other reactions}

\author{Qin-Song Zhou}\email{zhouqs@imu.edu.cn}
\affiliation{School of Physical Science and Technology, Inner Mongolia University, Hohhot 010021, China}
\affiliation{Research Center for Quantum Physics and Technologies, Inner Mongolia University, Hohhot 010021, China}
\affiliation{Inner Mongolia Key Laboratory of Microscale Physics and Atomic Manufacturing, Hohhot 010021, China}

\author{Zi-Yue Bai}\email{baizy15@lzu.edu.cn}
\affiliation{School of Physical Science and Technology, Lanzhou University, Lanzhou 730000, China}
\affiliation{Lanzhou Center for Theoretical Physics,
Key Laboratory of Theoretical Physics of Gansu Province,
Key Laboratory of Quantum Theory and Applications of MoE,
Gansu Provincial Research Center for Basic Disciplines of Quantum Physics, Lanzhou University, Lanzhou 730000, China}
\affiliation{Research Center for Hadron and CSR Physics, Lanzhou University and Institute of Modern Physics of CAS, Lanzhou 730000, China}

\author{Jun-Zhang Wang}\email{wangjzh@cqu.edu.cn}
\affiliation{Department of Physics, Chongqing University,
Chongqing 401331, China}

 \author{Hao Xu}\email{xuh2020@nwnu.edu.cn}
\affiliation{Institute of Theoretical Physics, College of Physics and Electronic Engineering,Northwest Normal University, Lanzhou 730070, China }

\author{Xiang Liu}\email{xiangliu@lzu.edu.cn}
\affiliation{School of Physical Science and Technology, Lanzhou University, Lanzhou 730000, China}
\affiliation{Lanzhou Center for Theoretical Physics,
Key Laboratory of Theoretical Physics of Gansu Province,
Key Laboratory of Quantum Theory and Applications of MoE,
Gansu Provincial Research Center for Basic Disciplines of Quantum Physics, Lanzhou University, Lanzhou 730000, China}
\affiliation{Research Center for Hadron and CSR Physics, Lanzhou University and Institute of Modern Physics of CAS, Lanzhou 730000, China}
\affiliation{MoE Frontiers Science Center for Rare Isotopes, Lanzhou University, Lanzhou 730000, China}

\date{\today}
\begin{abstract}
Recently, the BESIII Collaboration observed a $\rho$-like  structure $Y(2044)$ in $e^+e^-\to a_2(1320)\pi$, suggesting that $Y(2044)$ may be a candidate of vector meson $\rho(2D)$ by comparing resonance parameters. However, the theoretical prediction for the combined branching ratio $\Gamma_{e^+e^-}\mathcal{B}_{a_2(1320)\pi}$ for the pure $\rho(2D)$ state is about two orders of magnitude smaller than the experimental value.
To resolve this discrepancy and decipher the nature of $Y(2044)$, this work propose an $S$-$D$ mixing scheme to reanalyze the cross section of $e^+e^-\to a_2(1320)\pi$, and find that the aforementioned branching ratio discrepancy can be resolved.
Our results show that the $Y(2044)$ structure can be reproduced by introducing four theoretically predicted $S$-$D$ mixing $\rho$ meson states $\rho_{3S-2D}^{\prime}$, $\rho_{3S-2D}^{\prime\prime}$, $\rho_{4S-3D}^{\prime}$, and $\rho_{4S-3D}^{\prime\prime}$ as intermediate resonances, in which dominant contribution arises from $\rho_{3S-2D}^{\prime\prime}$ and their inference effect is also significant.
Furthermore, we reanalyzed five additional isospin vector processes $e^+e^-\to \omega\pi^0$, $e^+e^-\to f_1(1285)\pi^+\pi^-$, $e^+e^-\to \pi^+\pi^-$, $e^+e^-\to \rho \eta$, and $e^+e^-\to \eta^{\prime} \pi^+\pi^-$ based on the same  $S$-$D$ mixing framework, and simultaneously reproduced their experimental cross section data.
This work provides a unified framework to elucidate all observed $\rho$-like structures near 2 GeV in the $e^+e^-$ annihilation processes, and suggests that the $S$-$D$ mixing effect may be crucial for understanding the mass spectrum and decay behaviors of the higher $\rho$ meson states.
\end{abstract}

\maketitle

\section{Introduction}\label{sec1}

Light hadron spectroscopy remains a central focus of the particle physics community, offering a rich testing ground for understanding the nonperturbative regime of quantum chromodynamics (QCD). With the continuous accumulation of experimental data, a wealth of phenomena associated with light hadrons have been found. These discoveries not only contribute to the systematic construction of the conventional light hadron spectrum but also stimulate extensive discussions on exotic hadronic states, such as hybrids, glueballs and multiquark configurations. Such investigations are crucial for deepening our understanding of the hadron spectroscopy and the strong interaction.

The BESIII Collaboration has played a pivotal role in advancing light hadron spectroscopy~\cite{BESIII:2020nme}. Recently, BESIII reported precise measurements of the cross sections of the process $e^+e^- \to \eta\pi^+\pi^-$ at center-of-mass energies ranging from 2.00 to 3.08 GeV~\cite{BESIII:2023sbq}. Through a partial-wave analysis on $e^+ e^-\to \eta \pi^+ \pi^- $, a $\rho$-like structure, denoted as $Y(2044)$, was observed in the cross section distribution for the subprocess $e^+e^-\to a_2(1320)\pi$,  with a fitted mass $M = 2044 \pm 21 \pm 4$ MeV and a fitted width $\Gamma = 163 \pm 69 \pm 24$ MeV. Furthermore, the combined branching ratio $\Gamma_{e^+e^-}\mathcal{B}_{a_2(1320)\pi}$ was determined to be $(34.6 \pm 17.1 \pm 6.0)$ eV or $(137.1 \pm 73.3 \pm 2.1)$ eV, depending on whether constructive or destructive interference was assumed in the analysis of the cross section. The resonance parameters of $Y(2044)$ are consistent with theoretical expectations for the $\rho(2D)$ state, leading BESIII to propose $Y(2044)$ as a candidate for the $n^{2S+1}L_J = 2^3D_1$ state within the $\rho$ meson family.


However, this interpretation is challenged by a notable discrepancy: the theoretical prediction for di-leptonic width $\Gamma_{e^+e^-}$ and branching ratio $\mathcal{B}_{a_2(1320)\pi}$ of the pure $\rho(2D)$ state are $20$ eV~\cite{Wang:2021gle} and 3.6\%, respectively, whose combined product is at least nearly two orders of magnitude smaller than experimental value $(34.6 \pm 17.1 \pm 6.0)$ eV or $(137.1 \pm 73.3 \pm 2.1)$ eV~\cite{BESIII:2023sbq} reported for $Y(2044)$. In other word, this implies that if $Y(2044)$ is considered as pure $\rho(2D)$, its branching ratio for the decay $\rho(2D)\to a_2(1320)\pi$ would substantially exceed $100\%$. This inconsistency challenges the interpretation of $Y(2044)$ as a pure $\rho(2D)$ state and indicates the possible involvement of additional dynamics.
\begin{table}[htb]
  \centering
  \renewcommand\arraystretch{1.5}
  \caption{The resonance parameters and decay properties of excited $\rho$ meson states around 2 GeV. Here, the values of masses, total widths and di-leptonic widths are taken from the theoretical estimations in Ref.~\cite{Wang:2021gle}. The $\mathcal{B}_{a_2(1320)\pi}$ are estimated using the quark pair creation (QPC) model with the same model parameters as those employed in Ref.~\cite{Wang:2021gle}.}\label{T1}
  {\tabcolsep0.12in
  \begin{tabular}{ccccc}
  \toprule[1pt]
  \midrule[1pt]
  Parameters &  $\rho(3S)$ & $\rho(2D)$ & $\rho(4S)$ &  $\rho(3D)$ \\
  \midrule[1pt]
  Mass (MeV) & $1862$ & $2003$& $2180$ & $2283$ \\
  Total width (MeV) &  $115$ &  $179$ &  $102$ &  $158$ \\
  $\Gamma_{e^+e^-}$ (eV) &  $156$ &  $20$ &  $63$ &  $16$\\
  $\mathcal{B}_{a_2(1320)\pi}$ ($10^{-2}$) &  $13.5$ &  $3.6$ &  $9.6$ & $2.9$\\
\midrule[1pt]\bottomrule[1pt]
\end{tabular}
}
\end{table}
One possible explanation is the interference effect among several excited $\rho$ mesons. As shown in Table~\ref{T1}, four $\rho$ states are expected near 2 GeV, with relatively small mass gaps. Such proximity usually leads to strong interference effects in production and decay processes, potentially complicating the extraction of resonance parameters and decay characteristics of these intermediate $\rho$ meson states experimentally. Similar interference mechanisms have been successfully applied to describe other vector structures in the 2 GeV region~\cite{Wang:2021gle,Zhou:2022ark,Zhou:2022wwk,Liu:2022yrt,Wang:2020kte}. Motivated by this, we reanalyze the $e^+e^- \to a_2(1320)\pi$ cross section measured by BESIII~\cite{BESIII:2023sbq}, including four predicted $\rho$ meson states from Table~\ref{T1} as intermediate resonances. However, our analysis shows that the enhancement structure around 2044 MeV cannot be reproduced within the interference framework alone, whose details can be found in Sec.~\ref{sec2}.


Given that interference effects alone cannot account for the observed enhancement $Y(2044)$, an alternative explanation is required. It is well known that the di-leptonic width of a $D$-wave vector quarkonium is strongly suppressed compared to that of its $S$-wave counterpart, as predicted by quark potential models. In fact, this suppression mainly contributes to the large discrepancy between the theoretical and experimental combined branching ratios.  A potential mechanism of avoiding the di-leptonic rate suppression of a $D$-wave quarkonium is the $S$-$D$ mixing. Previous studies have demonstrated that $S$-$D$ mixing plays a crucial role in understanding the spectroscopy of meson families, including charmonium, bottomonium, and the $\omega$ meson~\cite{Wang:2019mhs,Wang:2022jxj,Wang:2023zxj,Peng:2024xui,Li:2021jjt,Bai:2022cfz,Li:2022leg,Liu:2024ets,Liu:2023gtx,Bai:2025knk}.

In Sec.~\ref{sec3}, we first perform a systematic analysis of the mass spectrum and decay behavior of $\rho$ mesons around 2 GeV within the $S$-$D$ mixing framework. In order to determine the $S$-$D$ mixing angles in these $\rho$ meson excitations, we need to identify some ideal decay processes, in which the observed enhancement structure is dominated by a single $S$-$D$ mixing state and the corresponding interference effects can be ignored. Based on this, we find two satisfactory processes $e^+e^-\to \omega\pi^0$~\cite{BESIII:2020xmw,Achasov:2016zvn} and $e^+e^-\to f_1(1285)\pi^+\pi^-$~\cite{BaBar:2007qju,BaBar:2022ahi} to fix the $S$-$D$ mixing angles between $\rho(3S)$ and  $\rho(2D)$, as well as between $\rho(4S)$ and  $\rho(3D)$, respectively.
Subsequently, we reanalyze the cross sections of the $e^+e^-\to a_2(1320)\pi$~\cite{BESIII:2023sbq} by considering the interference effect from four theoretically predicted $\rho$ meson states $\rho_{3S-2D}^{\prime}$, $\rho_{3S-2D}^{\prime\prime}$, $\rho_{4S-3D}^{\prime}$, and $\rho_{4S-3D}^{\prime\prime}$ as intermediate resonances and find that the enhancement structure associated with $Y(2044)$ can be reproduced well, which indicates that the problem of large discrepancy between theoretical and experimental combined branching ratios is fully solved. In addition, to further validate the reasonableness of the proposed mixing angles and to understand all  $\rho$-like structures observed in different $e^+e^-$ annihilation processes around 2 GeV,
we perform a combined fit to the cross sections of processes, $e^+e^-\to \omega\pi^0$~\cite{BESIII:2020xmw,Achasov:2016zvn}, $e^+e^-\to f_1(1285)\pi^+\pi^-$~\cite{BaBar:2007qju,BaBar:2022ahi}, $e^+e^-\to \pi^+\pi^-$~\cite{BaBar:2019kds}, $e^+e^-\to \rho \eta$~\cite{BESIII:2023sbq}, and $e^+e^-\to \eta^{\prime} \pi^+\pi^-$~\cite{BESIII:2020kpr} based on a consistent theoretical framework. Our results provide a unified understanding of all observed $\rho$-like enhancements near 2 GeV in the electron-positron annihilation processes.

\section{Data fitting of $e^+e^- \to a_2(1320)\pi$ in an interference framework without $S$-$D$ mixing}\label{sec2}

As mentioned in Sec.~\ref{sec1}, the small mass gaps among the $\rho$ mesons near 2 GeV, as listed in Table~\ref{T1}, may lead to significant interference effects in production processes. In this context, the resonance parameters and decay information of the enhancement structure in the cross section distribution, extracted using a single Breit-Wigner function, may deviate from theoretical predictions. Therefore, in this section, we attempt to understand $Y(2044)$ within the interference framework. Here, we will reanalyze the cross section data for $e^+e^-\to a_2(1320)\pi$ ~\cite{BESIII:2023sbq} by introducing the contributions from the four theoretically predicted pure $\rho$ meson states near 2 GeV as listed in Table~\ref{T1}.

The cross section of $e^+e^-\to a_2(1320)\pi$ can be modeled as the coherent sum of a continuum amplitude and a resonant amplitude~\cite{BESIII:2023sbq}
\begin{eqnarray}
\sigma(s)=|\mathcal{M}^{\rm{Dir}}+\sum_{k}e^{i\phi_{k}}\mathcal{M}_{k}|^2,
\end{eqnarray}
where $\phi_{k}$ is the relative phase angle between the different amplitudes. The continuum amplitude $\mathcal{M}^{\rm{Dir}}$ is written as
\begin{eqnarray}\label{eq1}
\mathcal{M}^{\rm{Dir}}=C_0 \cdot s^{-n}\sqrt{\Phi_2(s)},
\end{eqnarray}
where $\Phi_2(s)$ is two-body phase space. The resonant amplitude $\mathcal{M}_{k}$ is described with a Breit-Wigner function as
\begin{eqnarray}\label{eq2}
\mathcal{M}_{k}=\frac{\sqrt{12\pi\Gamma_{e^+e^-}^{\rho^*_k}\mathcal{B}_{a_2(1320)\pi}^{\rho^*_k}\Gamma_{\rho^*_k}^{\rm{tot}}}}{s-M_{\rho^*_k}^2+iM_{\rho^*_k}\Gamma_{\rho^*_k}^{\rm{tot}}}\sqrt{\frac{\Phi_2(s)}{\Phi_2(M_{\rho^*_k}^2)}},
\end{eqnarray}
where $M_{\rho^*}$, $\Gamma_{\rho^*}^{\rm{tot}}$, $\Gamma^{\rho^*}_{e^+e^-}$, and $\mathcal{B}_{a_2(1320)\pi}^{\rho^*}$ are the mass, total width, partial width to $e^+e^-$, and branching ratio for the decay $\rho^*\to a_2(1320) \pi$ of the intermediate excited $\rho$ mesons, respectively. These values are taken from the theoretical calculations presented in Ref.~\cite{Wang:2021gle}, as summarized in Table \ref{T1}. Here, the values of masses, total widths and di-leptonic widths are taken from the theoretical estimations in Ref.~\cite{Wang:2021gle}. The $\mathcal{B}_{a_2(1320)\pi}$ are estimated using the quark pair creation (QPC) model with the same model parameters as those employed in Ref.~\cite{Wang:2021gle}. The interference phase angles and parameters included in continuum amplitude are treated as free parameters and determined by fitting the cross section data of $e^+e^-\to a_2(1320)\pi$~\cite{BESIII:2023sbq}. The fitting results with these parameters are listed in Table~\ref{T2}.

\begin{table}[htb]
  \centering
  \renewcommand\arraystretch{1.5}
  \caption{The parameters obtained by fitting the cross section of $e^+e^-\to a_2(1320)\pi$ measured by the BESIII Collaboration~\cite{BESIII:2023sbq}.}\label{T2}
   {\tabcolsep0.11in
  \begin{tabular}{ccccccc}
  \toprule[1pt]
  \midrule[1pt]
  Parameters      & Values          & Parameters     & Values \\
  \midrule[1pt]

  $n$    &  $1.57\pm0.01$   & $C_0$         & $481.96\pm8.65$\\

  $\phi_1$ (rad) &  $5.75\pm0.08$   & $\phi_2$ (rad)  & $0.38\pm0.24$\\

  $\phi_3$ (rad) &  $3.44\pm0.09$   & $\phi_4$ (rad)  & $2.92\pm0.43$\\
\midrule[1pt]
\bottomrule[1pt]
\end{tabular}
}
\end{table}

\begin{figure}[!htbp]
  \centering
  \begin{tabular}{c}
  \includegraphics[width=200pt]{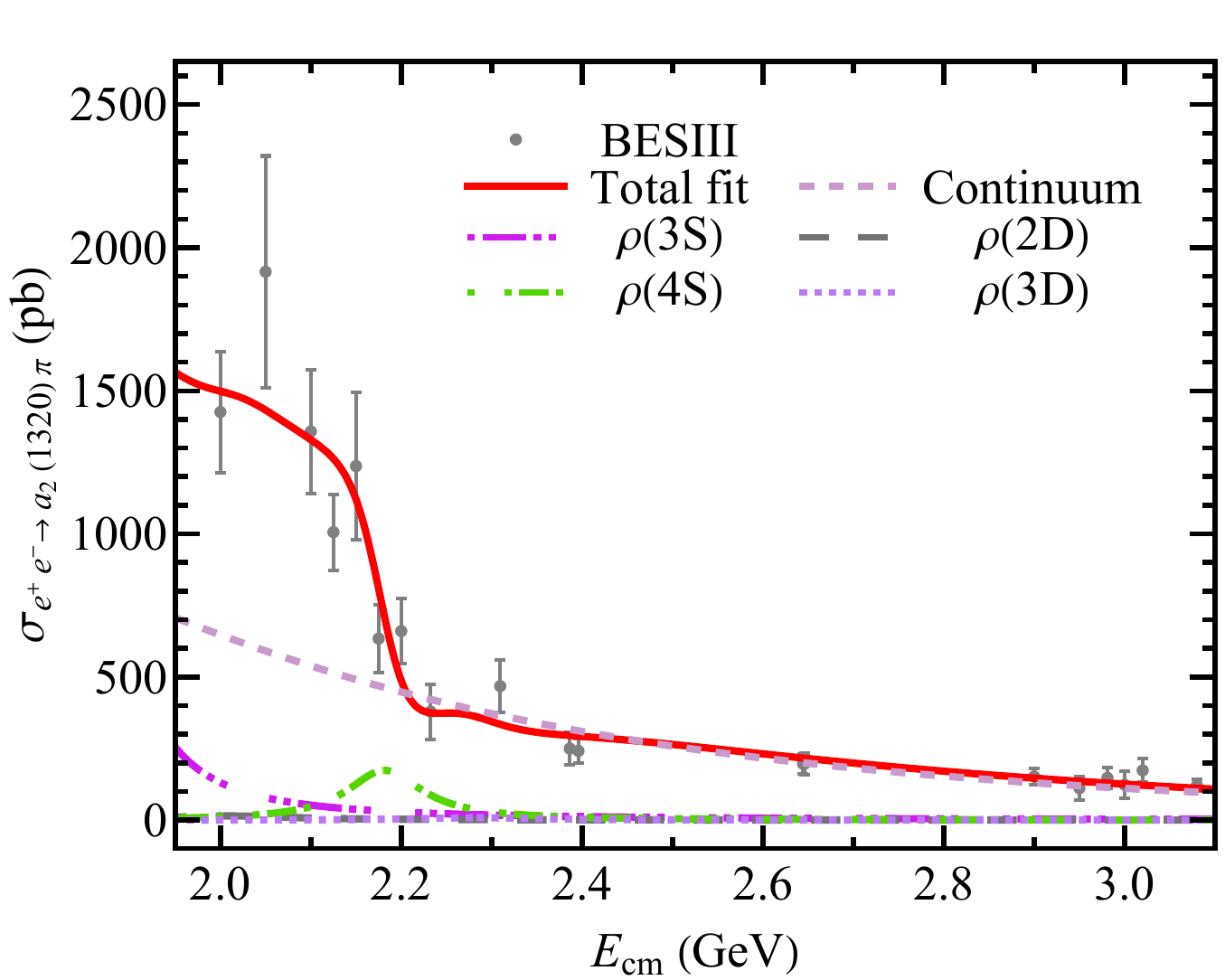}
  \end{tabular}
  \caption{Fit to the cross section of the process $e^+e^-\to a_2(1320)\pi$ measured by the BESIII Collaboration~\cite{BESIII:2023sbq}.}\label{F1}
\end{figure}

Fig.~\ref{F1} presents the fitting result for the cross section of $e^+e^-\to a_2(1320)\pi$~\cite{BESIII:2023sbq} with a fit quality of $\chi^2/\rm{n.d.f.}=1.33$, where the gray solid dot with error bars are BESIII data, the red solid curve is the total fitting result, the short dashed line is the continuum contribution, and the purple dot-dot-dashed line, the gray dashed line, the green dot-dashed line, the dot line correspond to the $\rho(3S)$, $\rho(2D)$, $\rho(4S)$, $\rho(3D)$ contributions, respectively. As shown in Fig. \ref{F1}, the peak structure near 2044 MeV cannot be reproduced by considering the interference effects among the four pure four excited $\rho$ meson states near 2 GeV listed in Table~\ref{T1}. This implies that the unsuccessful fit may be due to the limited contribution of $\rho(2D)$ (with a mass of around 2044 MeV) to the $e^+e^-\to a_2(1320)\pi$ process.

\section{Data fitting of six isospin vector $e^+e^-$ annihilation processes in an interference framework with $S$-$D$ mixing}\label{sec3}

\subsection{Spectra and decay properties of excited $\rho$ meson states near 2 GeV including $S$-$D$ mixing}\label{sec3A}

\begin{figure*}[!htbp]
  \centering
  \begin{tabular}{ccc}
  \subfigure{\label{F2a}\includegraphics[width=200pt]{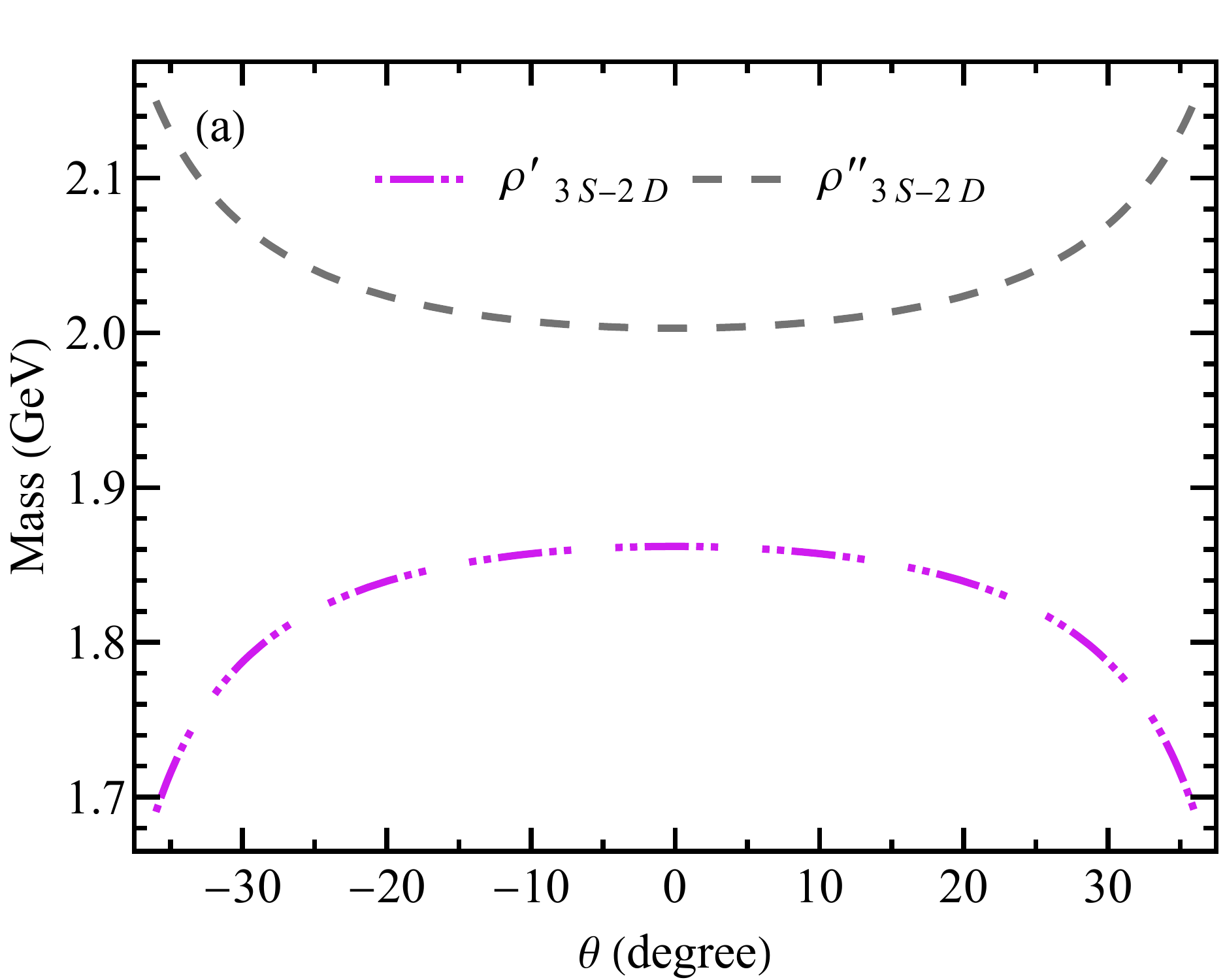}}&$\quad$&\subfigure{\label{F2b}\includegraphics[width=200pt]{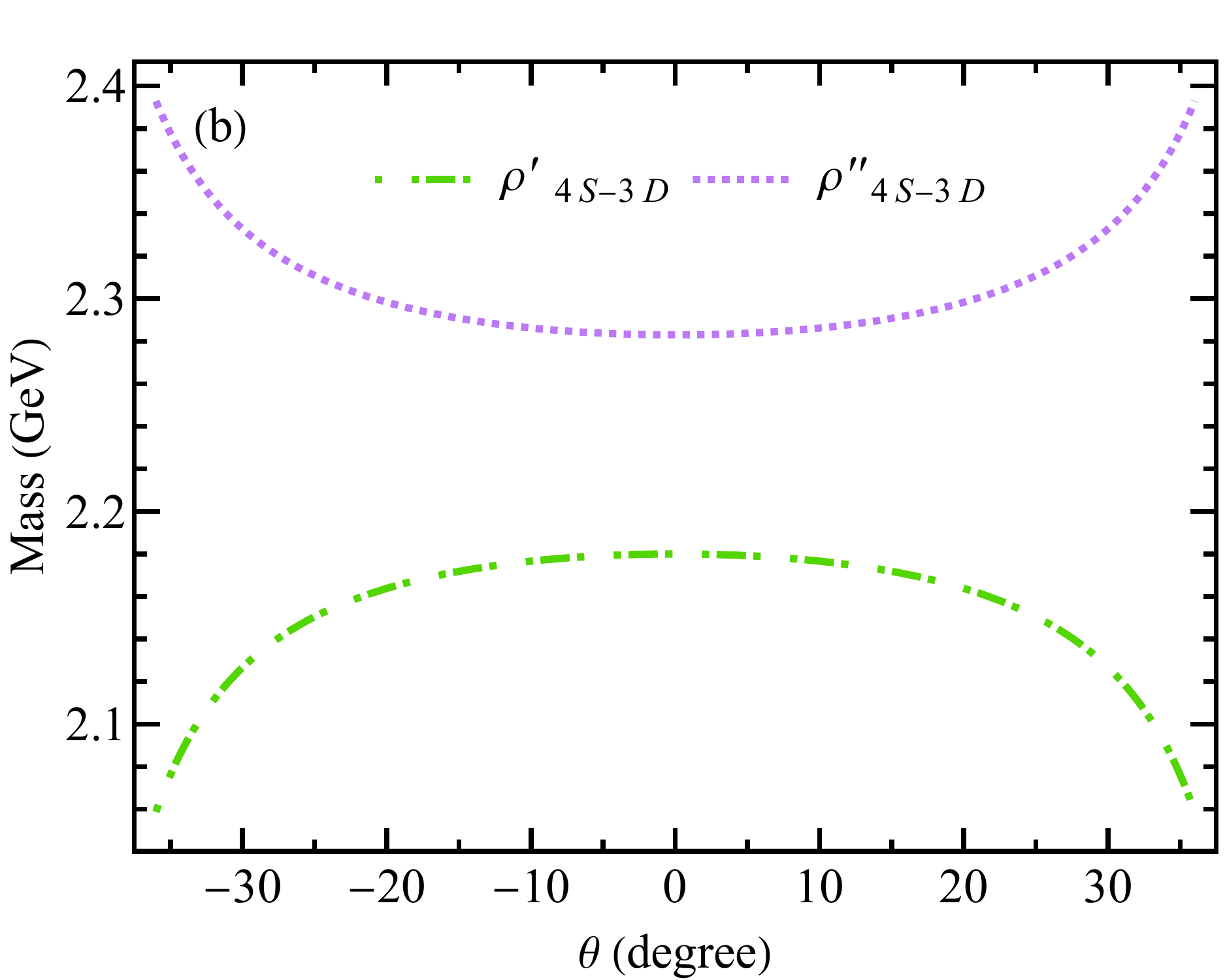}}
  \end{tabular}
  \caption{The dependence of masses of mixed states on mixing angele $\theta$. The sub-figure (a) shows the masses of mixed states $\rho_{3S-2D}^{\prime}$ and $\rho_{3S-2D}^{\prime\prime}$ as a function of the mixing angle, while sub-figure (b) shows the masses of mixed states $\rho_{4S-3D}^{\prime}$ and $\rho_{4S-3D}^{\prime\prime}$  as a function of the mixing angle.}\label{F2}
\end{figure*}

In this subsection, we will investigate the mass spectrum and decay characteristics of the excited $\rho$ mesons near 2 GeV under the $S$-$D$ mixing framework.
Within the $S$-$D$ mixing framework, the $\rho(nS)$-$\rho((n-1)D)$ mixing can be expressed as~\cite{Bai:2025knk}
\begin{eqnarray}\label{mix}
\begin{pmatrix}
|\rho_{nS-(n-1)D}^\prime\rangle \\
|\rho_{nS-(n-1)D}^{\prime\prime}\rangle
\end{pmatrix}=
\begin{pmatrix}
\cos\theta &\sin\theta\\
-\sin\theta&\cos\theta
\end{pmatrix}
\begin{pmatrix}
|\rho(nS)\rangle\\
|\rho((n-1)D)\rangle
\end{pmatrix},
\end{eqnarray}
where $\theta$ denotes the mixing angle between $\rho(nS)$ and $\rho((n-1)D)$. The masses of $\rho_{nS-(n-1)D}^\prime$ and $\rho_{nS-(n-1)D}^{\prime\prime}$ can be determined by the masses of $m_{nS}$, $m_{(n-1)D}$, and the mixing angle $\theta$, i.e.~\cite{Bai:2025knk},
\begin{eqnarray}\nonumber
m_{\rho_{nS-(n-1)D}^\prime}^2&=&\frac{1}{2}\bigg(m_{\rho(nS)}^2+m_{\rho((n-1)D)}^2\\ \label{eq5}
&&-\sqrt{(m_{\rho(nS)}^2-m_{\rho((n-1)D)}^2)^2\sec^2{2\theta}}\bigg),\\ \nonumber
\\ \nonumber
m_{\rho_{nS-(n-1)D}^{\prime\prime}}^2&=&\frac{1}{2}\bigg(m_{\rho(nS)}^2+m_{\rho((n-1)D)}^2\\ \label{eq6}
&&+\sqrt{(m_{\rho(nS)}^2-m_{\rho((n-1)D)}^2)^2\sec^2{2\theta}}\bigg),
\end{eqnarray}
where the $m_{nS}$ and $m_{(n-1)D}$ are the masses of pure $\rho(nS)$ and $\rho((n-1)D)$, respectively, calculated by an unquenched relativized potential model~\cite{Wang:2021gle}.
By substituting the masses of the pure $\rho$ mesons from Table~\ref{T1} into Eqs. (\ref{eq5}) and (\ref{eq6}), we can derive the dependence of the masses of mixed states on the mixing angle $\theta$, as shown in Fig.~\ref{F2}. This figure indicates that the mixed state mass $m_{\rho_{nS-(n-1)D}^{\prime}}$ decreases with increasing absolute value of the mixing angle $\theta$,whereas $m_{\rho_{nS-(n-1)D}^{\prime\prime}}$ increases as the absolute value of $\theta$ increases.
Thus, the $\rho$-like structure with a mass of about 2044 MeV mentioned in Sec.~\ref{sec2} may be one of the mixed states $\rho_{3S-2D}^{\prime\prime}$ derived from $\rho(3S)$ and $\rho(2D)$.
The determination of mixing angles will be discussed later in this subsection.

Due to the distortion caused by the unavoidable interference effects in production processes, it is necessary to identify ideal processes to determine the $S$-$D$ mixing angles between $\rho(3S)$ and $\rho(2D)$ ($\theta_{3S-2D}$) and between $\rho(4S)$ and $\rho(3D)$ ($\theta_{4S-3D}$). To achieve this goal, we will investigate the decay properties of the $\rho$ meson states around 2 GeV within the $S$-$D$ mixing framework. Here, we employ the quark pair creation (QPC) model \cite{Micu:1968mk,LeYaouanc:1977gm,Jacob:1959at} to study the Okubo-Zweig-Iizuka (OZI) allowed two-body strong decays of these mixed $\rho$ mesonic states. The partial widths of the $\rho_{nS-(n-1)D}^{\prime}$ and $\rho_{nS-(n-1)D}^{\prime\prime}$ decay to $B$ and $C$ can be expressed as
\begin{eqnarray}
\Gamma_{\rho_{nS-(n-1)D}^{\prime}\to BC} &= & \frac{\pi}{4} \frac{|\mathbf{P}|}{m_{\rho_{nS-(n-1)D}^\prime}^2} \sum_{JL} \left|\mathcal{M}_{\rho_{nS-(n-1)D}^{\prime}}^{JL}(\mathbf{P})\right|^2,\\ \nonumber
\\
\Gamma_{\rho_{nS-(n-1)D}^{\prime\prime}\to BC} &=& \frac{\pi}{4} \frac{|\mathbf{P}|}{m_{\rho_{nS-(n-1)D}^{\prime\prime}}^2} \sum_{JL} \left|\mathcal{M}_{\rho_{nS-(n-1)D}^{\prime\prime}}^{JL}(\mathbf{P})\right|^2,
\end{eqnarray}
where $\mathbf{P}=\mathbf{P}_B=-\mathbf{P}_C$ is the three-momentum of the particle $B$ in the center-of-mass frame, $\textbf{\textit{L}}$ and $\textbf{\textit{J}}$ denote the relative orbital angular and total spin momentum between final states $B$ and $C$. The mixed partial wave amplitudes of $\mathcal{M}_{\rho_{nS-(n-1)D}^{\prime}}^{JL}(\mathbf{P})$ and $\mathcal{M}_{\rho_{nS-(n-1)D}^{\prime\prime}}^{JL}(\mathbf{P})$ can be represented as a combination of the pure state amplitudes $\mathcal{M}_{\rho(nS)}^{JL}(\mathbf{P})$ and $\mathcal{M}_{\rho((n-1)D)}^{JL}(\mathbf{P})$, and are denoted by
\begin{eqnarray}
\begin{pmatrix}
\mathcal{M}_{\rho_{nS-(n-1)D}^{\prime}}^{JL}(\mathbf{P})\\
\\
\mathcal{M}_{\rho_{nS-(n-1)D}^{\prime\prime}}^{JL}(\mathbf{P})
\end{pmatrix}=
\begin{pmatrix}
\cos \theta & \sin \theta\\
\\
-\sin\theta & \cos \theta
\end{pmatrix}
\begin{pmatrix}
\mathcal{M}_{\rho(nS)}^{JL}(\mathbf{P})\\
\\
\mathcal{M}_{\rho((n-1)D)}^{JL}(\mathbf{P})
\end{pmatrix},
\end{eqnarray}
where the pure state amplitudes $\mathcal{M}_{\rho(nS)}^{JL}(\mathbf{P})$ and $\mathcal{M}_{\rho((n-1)D)}^{JL}(\mathbf{P})$ adopt the results of Refs. \cite{Wang:2021gle,Wang:2021abg}.

The di-leptonic widths of the mixed $\rho_{nS-(n-1)D}^\prime$ and $\rho_{nS-(n-1)D}^{\prime\prime}$ are given by~\cite{Wang:2021gle,Wang:2021abg,Godfrey:1985xj,Zhou:2022ark,Zhou:2022wwk}
\begin{eqnarray}\label{Gee}
\Gamma^{e^+e^-}_{\rho_{nS-(n-1)D}^\prime} &=& \frac{4\pi\alpha^2 m_{\rho_{nS-(n-1)D}^\prime}}{3}  \left| \mathcal{M}_{\rho_{nS-(n-1)D}^\prime}^{e^+e^-} \right|^2, \\
\Gamma^{e^+e^-}_{\rho_{nS-(n-1)D}^{\prime\prime}} &=& \frac{4\pi\alpha^2 m_{\rho_{nS-(n-1)D}^{\prime\prime}}}{3}  \left|  \mathcal{M}_{\rho_{nS-(n-1)D}^{\prime\prime}}^{e^+e^-}  \right|^2.
\end{eqnarray}
Similar to the strong decay, the di-leptonic decay amplitudes can also be expressed as a function of the pure-state di-leptonic decay amplitudes as follows
\begin{eqnarray}
\begin{pmatrix}
\mathcal{M}_{\rho_{nS-(n-1)D}^{\prime}}^{e^+e^-}\\
\\
\mathcal{M}_{\rho_{nS-(n-1)D}^{\prime\prime}}^{e^+e^-}
\end{pmatrix}=
\begin{pmatrix}
\cos \theta & \sin \theta\\
\\
-\sin\theta & \cos \theta
\end{pmatrix}
\begin{pmatrix}
\mathcal{M}_{\rho(nS)}^{e^+e^-}\\
\\
\mathcal{M}_{\rho((n-1)D)}^{e^+e^-}
\end{pmatrix},
\end{eqnarray}
where the di-leptonic decay amplitudes of the pure higher $\rho$ meson states can be related to the zero-point behavior of their radial wave functions, as specified by the formula detailed in Refs.~\cite{Wang:2021gle,Wang:2021abg,Godfrey:1985xj}.

\begin{figure*}[!htbp]
  \centering
  \begin{tabular}{ccc}
  \subfigure{\label{F3a}\includegraphics[width=200pt]{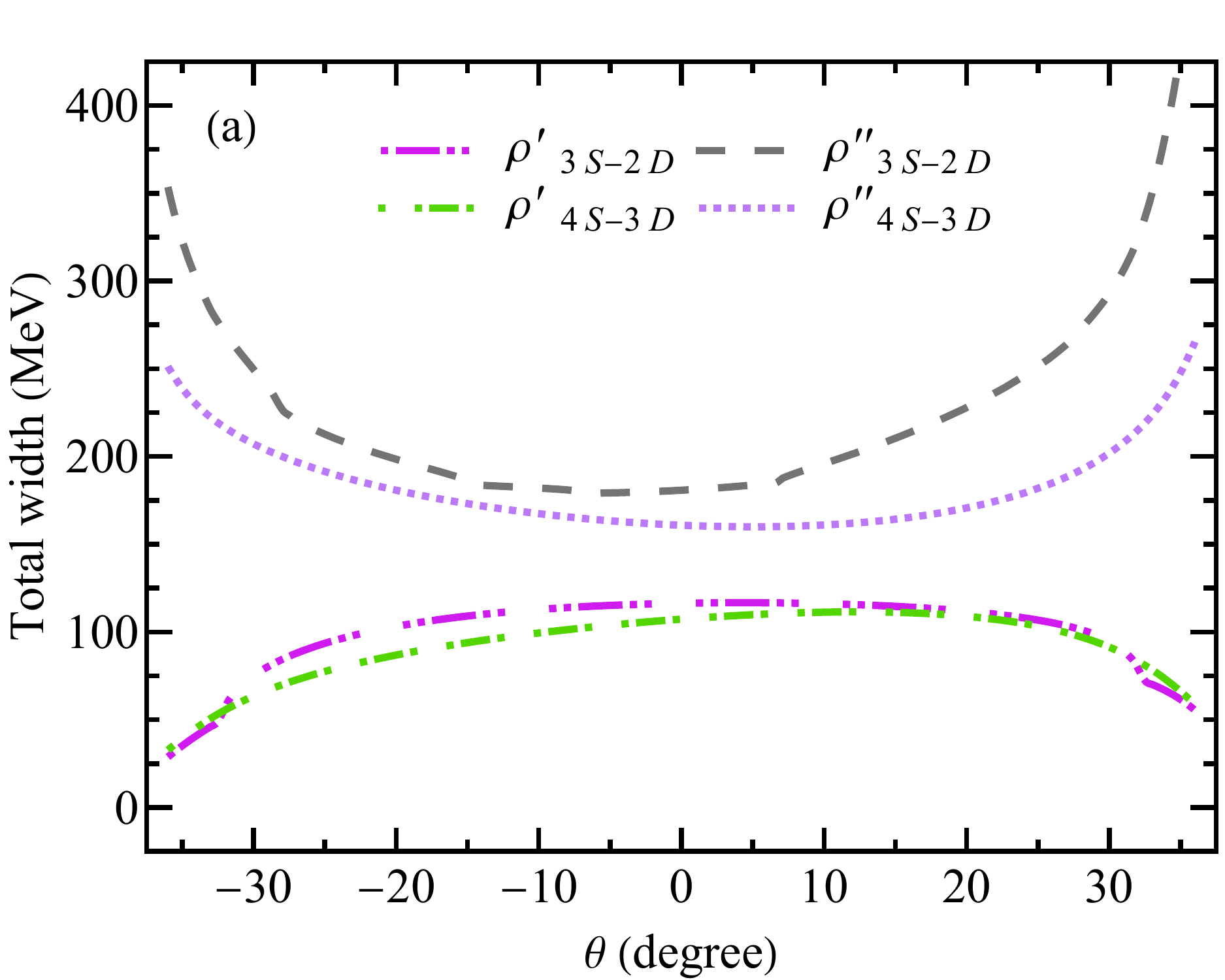}}&$\quad$&\subfigure{\label{F3b}\includegraphics[width=200pt]{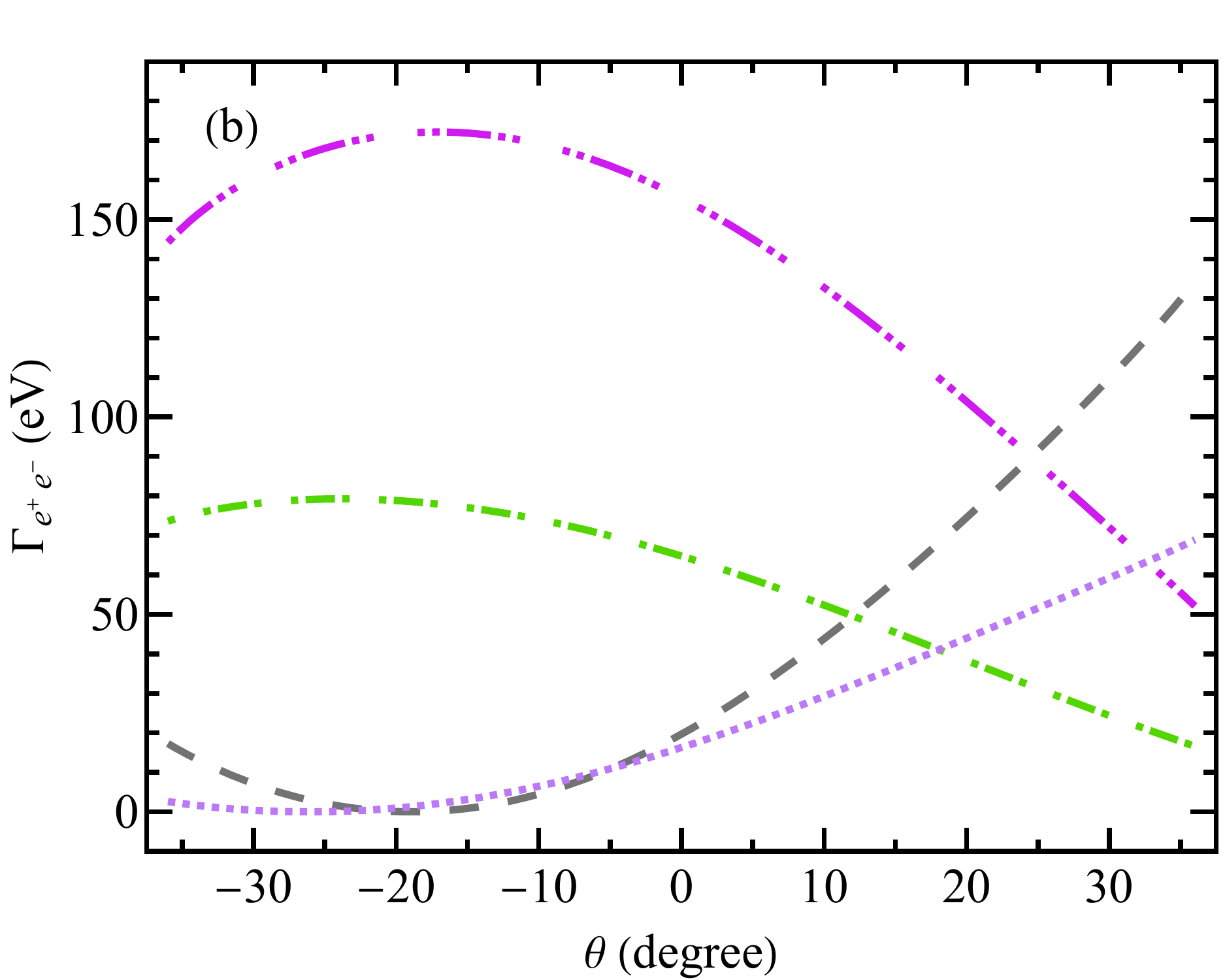}}\\
  \subfigure{\label{F3c}\includegraphics[width=200pt]{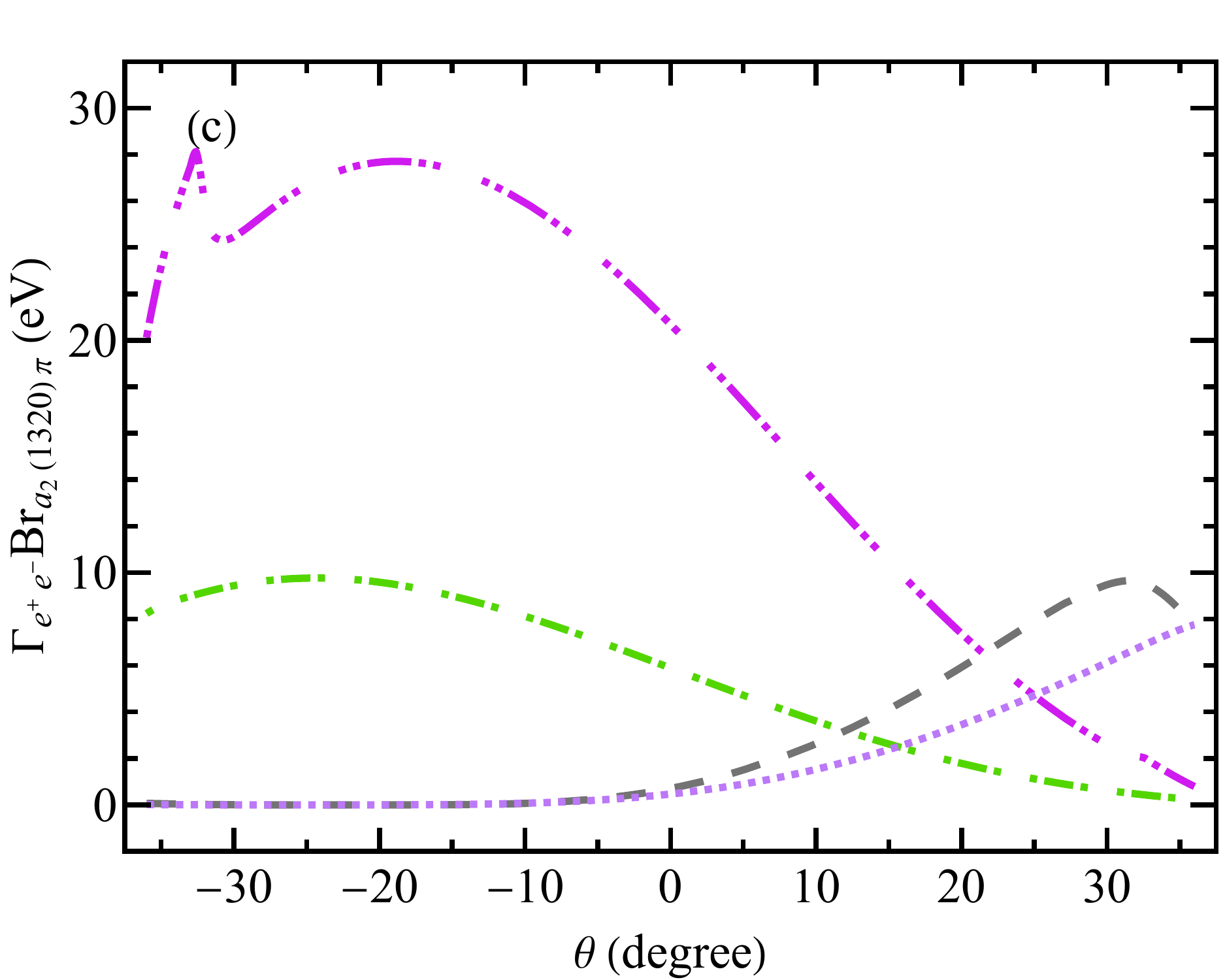}}&$\quad$&\subfigure{\label{F3d}\includegraphics[width=200pt]{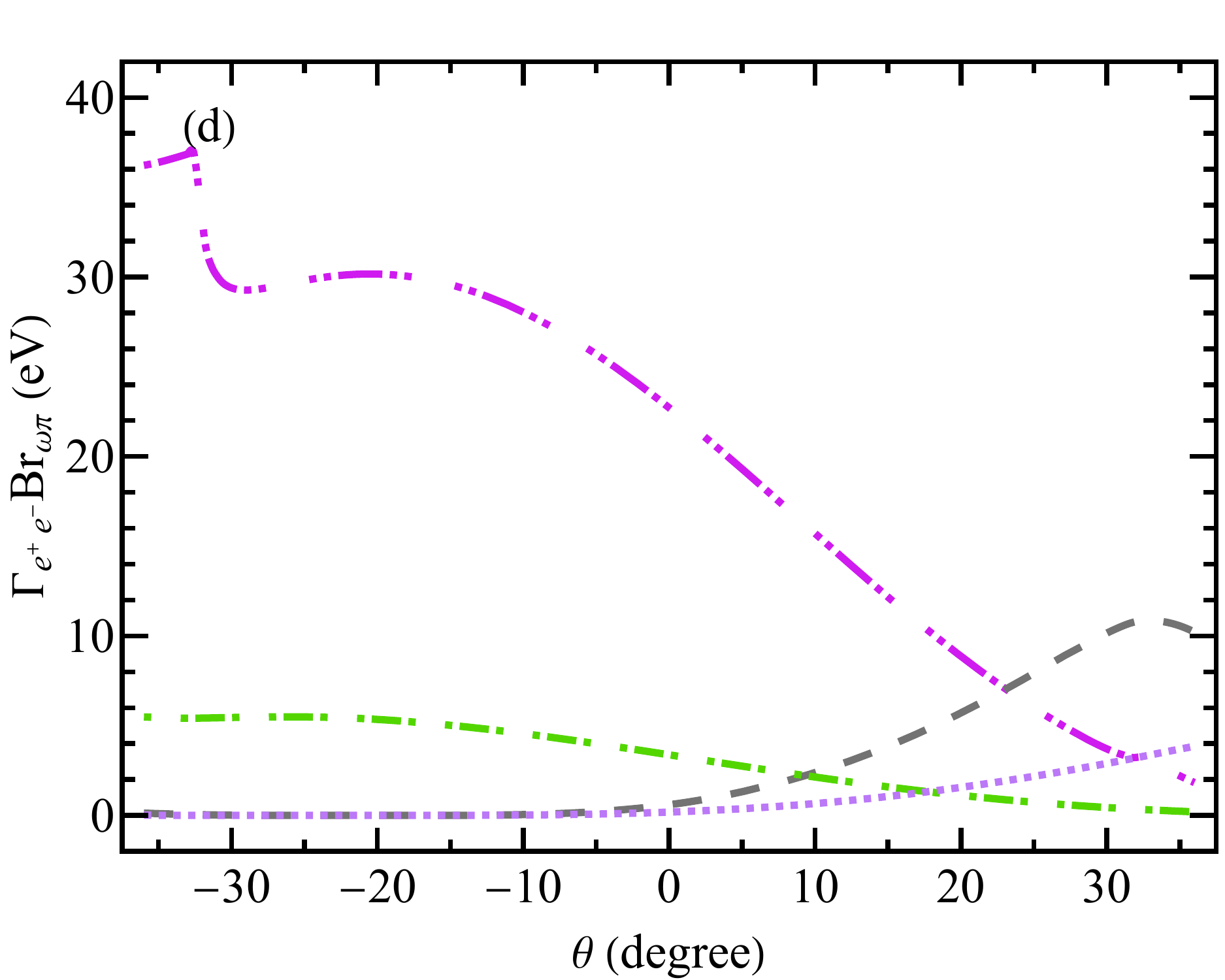}}\\
  \subfigure{\label{F3e}\includegraphics[width=200pt]{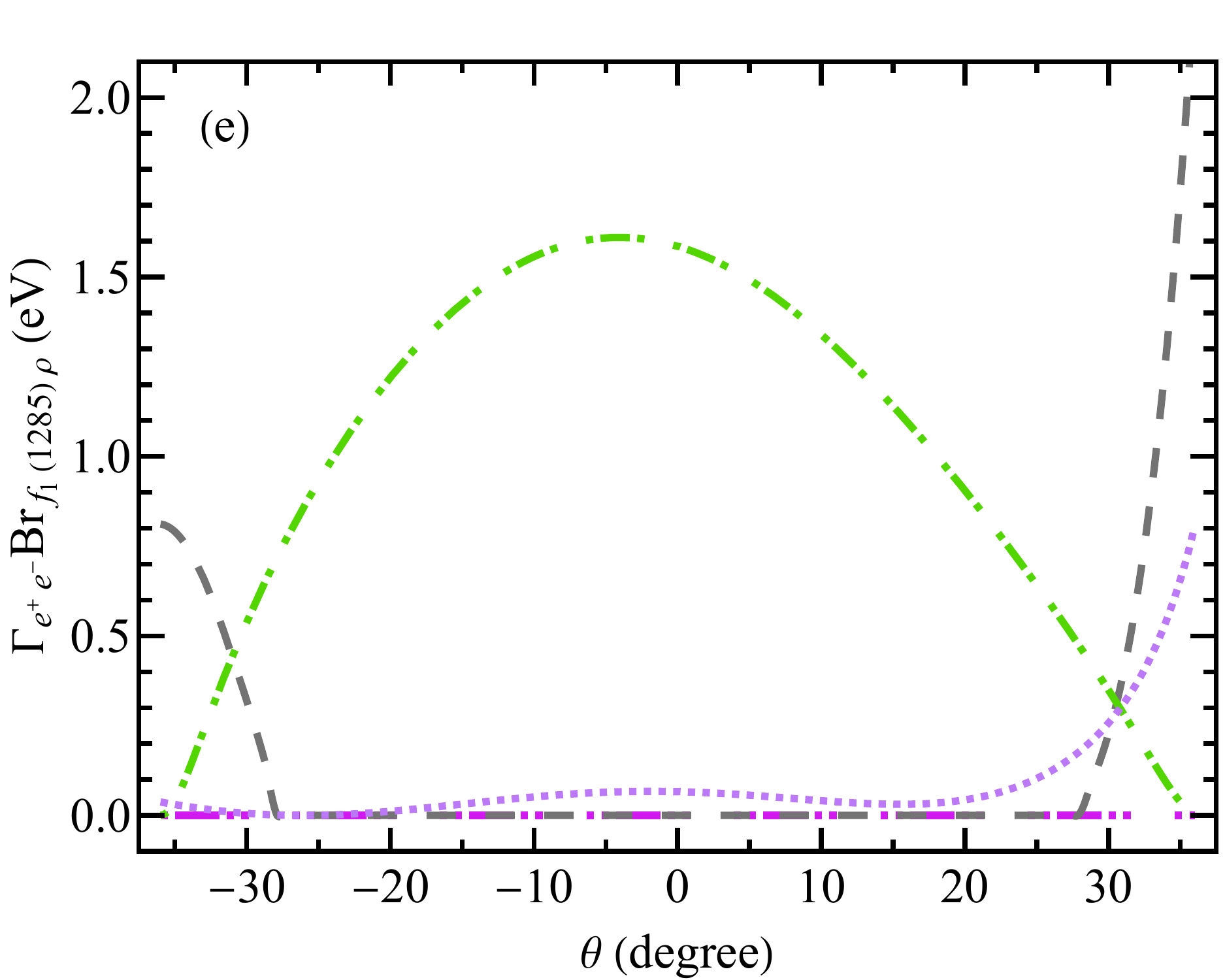}}&$\quad$&\subfigure{\label{F3f}\includegraphics[width=200pt]{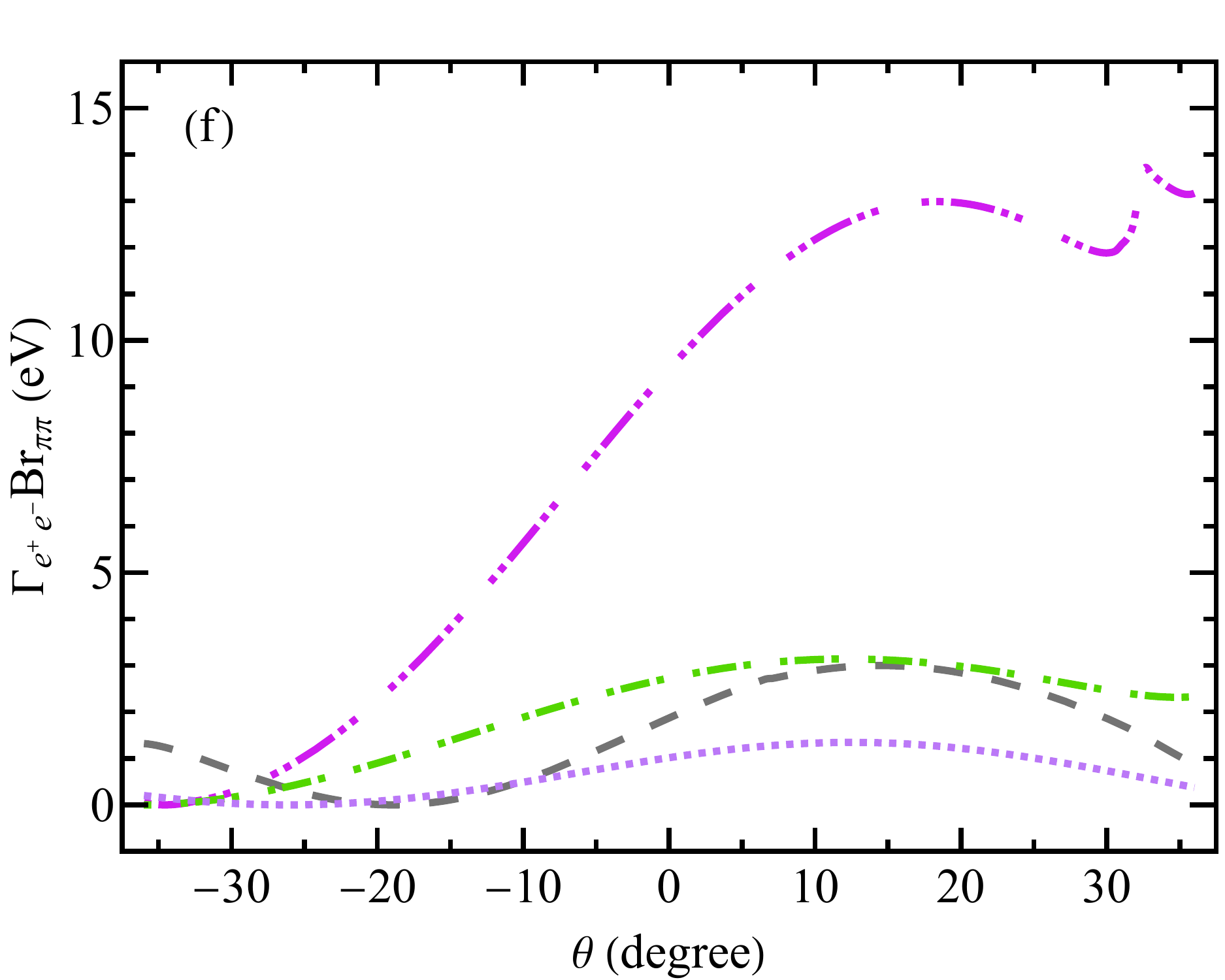}}\\
  \subfigure{\label{F3g}\includegraphics[width=200pt]{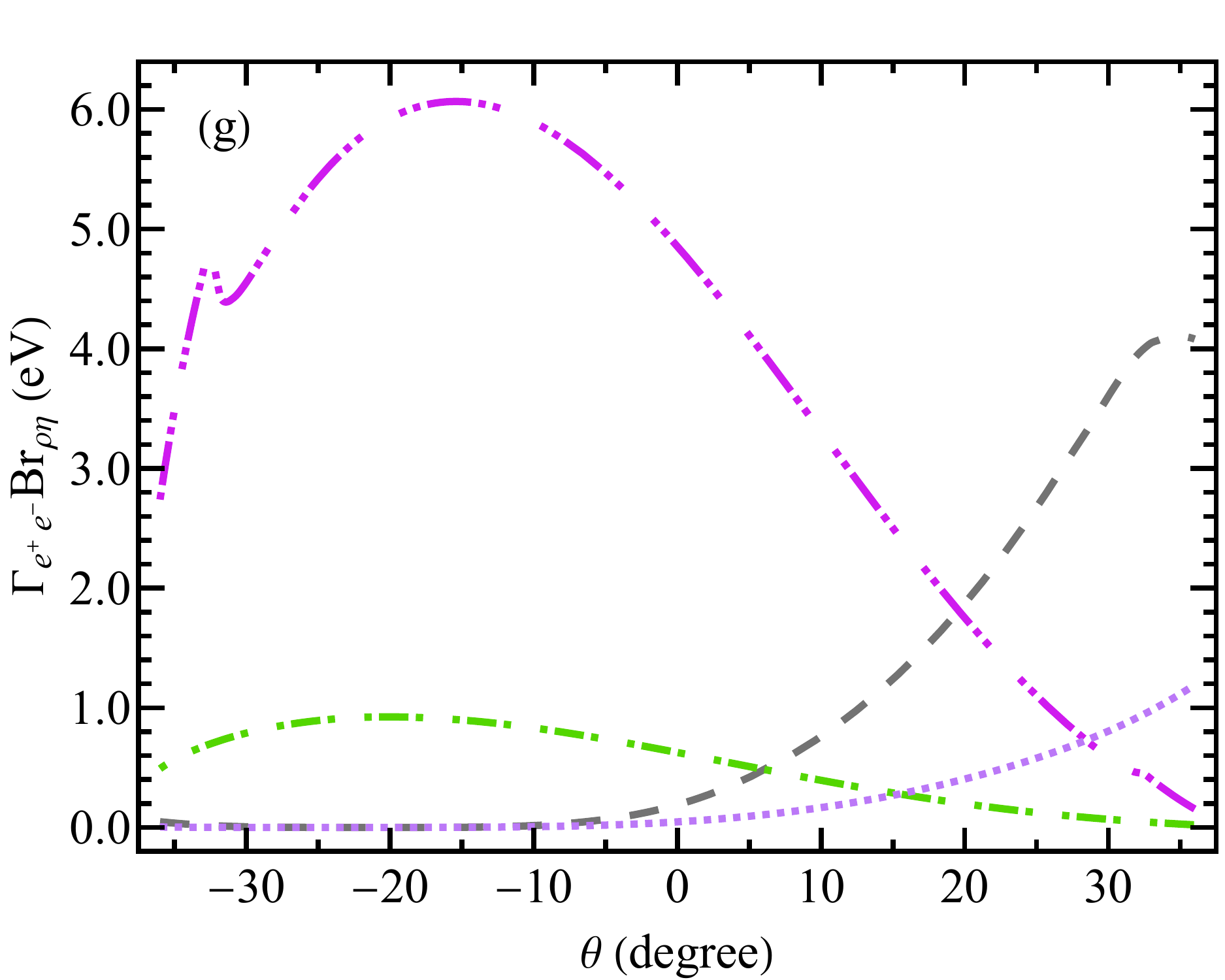}}&$\quad$&\subfigure{\label{F3h}\includegraphics[width=200pt]{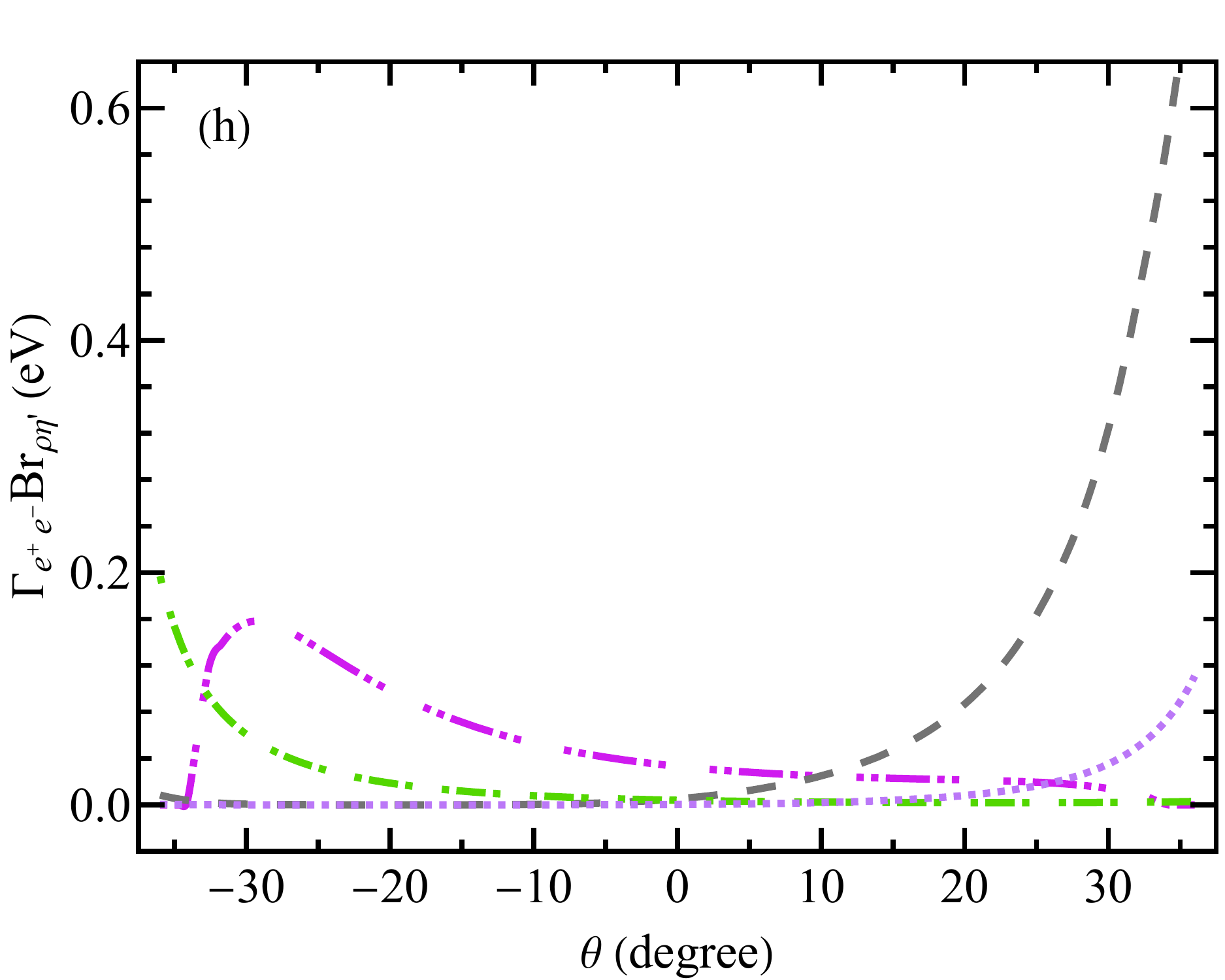}}\\
  \end{tabular}
  \caption{ The decay behavior of $\rho_{3S-2D}^{\prime}$, $\rho_{3S-2D}^{\prime\prime}$, $\rho_{4S-3D}^{\prime}$, and $\rho_{4S-3D}^{\prime\prime}$ as functions of the mixing angle $\theta$.}\label{F3}
\end{figure*}

In Fig.~\ref{F3}, we present the decay behaviors of $\rho_{3S-2D}^{\prime}$, $\rho_{3S-2D}^{\prime\prime}$, $\rho_{4S-3D}^{\prime}$, and $\rho_{4S-3D}^{\prime\prime}$ as functions of the mixing angle $\theta$. Fig.~\ref{F3a} presents the dependence of total widths of the mixed $\rho$ meson states on the mixing angle, with line shapes closely resemble those of the dependence of masses on the mixing angle shown in Fig.~\ref{F2}. This indicates that the variations in their total width are primarily influenced by phase space. Fig.~\ref{F3b} presents the dependence of the di-lepton widths of the mixed $\rho$ meson states on the mixing angle, showing that the di-leptonc widths of $\rho_{3S-2D}^{\prime}$ and $\rho_{4S-3D}^{\prime}$ decrease with increasing positive mixing angle, whereas those of $\rho_{3S-2D}^{\prime\prime}$ and $\rho_{4S-3D}^{\prime\prime}$ increase.
These six channels illustrated in Figs.~\ref{F3c}-\ref{F3h} encompass all isospin vector processes studied to date in electron-positron annihilation~\cite{ParticleDataGroup:2024cfk}. From Fig.~\ref{F3}, we can see that the impact of $S$-$D$ mixing effects on the decay behaviors of these $\rho$ meson states is highly significant. Next, we need to identify ideal processes to  determine the values of mixing angles $\theta_{3S-2D}$ and $\theta_{4S-3D}$. Here, these ideal processes are defined as those in which a specific intermediate meson $\rho$ meson state dominates over other $\rho$ meson states within the 2 GeV energy range and the corresponding interference effect can be ignored. The logic for determining these two mixing angles is as follows:
\begin{itemize}
\item[(1)] Fig.~\ref{F3c} indicates that if the $\rho_{3S-2D}^{\prime\prime}$ contributes significantly to $e^+e^-\to a_2(1320)\pi$, the mixing angle between $\rho(3S)$ and $\rho(2D)$ must satisfy $\theta_{3S-2D}>0$.
\item[(2)] Fig.~\ref{F3d} shows that if the mixing angle $\theta_{4S-3D}>0$ is considered, the combined branching ratios of $\rho_{4S-3D}^{\prime}$ to $\omega\pi$ are relatively small and the contribution of $\rho_{3S-2D}^{\prime\prime}$ to the structure $Y(2034)$ observed in the process $e^+e^-\to \omega\pi^0$ ~\cite{BESIII:2020xmw} will be the most significant. It is noteworthy that although $\rho_{3S-2D}^{\prime}$ and $\rho_{4S-3D}^{\prime\prime}$ may also have significant contributions to $e^+e^-\to \omega\pi^0$, but their masses are considerably distant from 2034 MeV, thus the interference contributions from $\rho_{3S-2D}^{\prime}$ and $\rho_{4S-3D}^{\prime\prime}$ are unlikely significant for producing the $Y(2034)$ structure.
    In this context, this resonance structure can be regarded as a good candidate for $\rho_{3S-2D}^{\prime\prime}$.
    Therefore, substituting the mass of $Y(2034)$ measured by BESIII Collaboration~\cite{BESIII:2020xmw} into Eq.~(\ref{eq6}), the central value of $\theta_{3S-2D}=\pm23.4^{\circ}$ is obtained.
\item[(3)] Fig.~\ref{F3e} shows that $e^+e^-\to f_1(1285)\rho$ is a ideal process to determine the the mixing angle $\theta_{4S-3D}$. The study carried out by the \textit{BABAR} Collaboration on process $e^+e^-\to f_1(1285)\pi^+\pi^-$~\cite{BaBar:2022ahi} reveals that $\pi^+\pi^-$ is dominantly from $\rho(770)$. When $\rho_{3S-2D}^{\prime}$ and $\rho_{3S-2D}^{\prime\prime}$ serve as intermediate resonant states in $e^+e^-\to f_1(1285)\pi^+\pi^-$, their contributions will be suppressed because $\rho(770)$ is off-shell.
    Additionally, the \textit{BABAR} Collaboration observed an enhancement structure in $e^+e^-\to f_1(1285)\pi^+\pi^-$~\cite{BaBar:2022ahi} processes only around 2150 MeV, with no evidence of other enhancement structure at higher energy regions, and this indicates that the contribution of the $\rho_{4S-3D}^{\prime\prime}$ in this process should be not significant and the corresponding mixing angle $\theta_{4S-3D}$ is also unlikely to be very large. Therefore, we have sufficient justification to identify the $\rho$-like state with a mass of $2150\pm40\pm50$ MeV observed in process $e^+e^-\to f_1(1285)\pi^+\pi^-$~\cite{BaBar:2022ahi} as the $\rho_{4S-3D}^{\prime}$.
    According to Eq.~(\ref{eq5}), the central value of $\theta_{4S-3D}$ is determined to be $\theta_{4S-3D} = \pm25.1^{\circ}$. Here, the concrete sign will be determined in sub-section~\ref{Sec3B} through fitting a series of cross section data of electron-positron annihilation processes.
\end{itemize}

Fig.~\ref{F3f} indicates that when both $\theta_{3S-2D}$ and $\theta_{4S-3D}$ are greater than 0, the combined branching ratio $\Gamma_{e^+e^-}\mathcal{B}_{\pi\pi}$ of $\rho_{3S-2D}$ is significantly larger than those of the other three $\rho$ meson states. Therefore, the $e^+e^-\to \pi^+\pi^-$ can serves as a crucial process to determine the mixing angle $\theta_{3S-2D}$. Unfortunately, there are no experimental measurements of cross sections for $e^+e^-\to \pi^+\pi^-$ processes below 2 GeV.
The combined branching ratios $\Gamma_{e^+e^-}\mathcal{B}_{\rho\eta}$ and $\Gamma_{e^+e^-}\mathcal{B}_{\rho\eta^{\prime}}$ as functions of the mixing angle presented in Fig.~\ref{F3g} and ~\ref{F3h},respectively, suggest that $e^+e^-\to\rho\eta$ and $e^+e^-\to\rho\eta^{\prime}$ may be more suitable for determining $\theta_{3S-2D}$.
However, due to the poor precision and statistical significance of the experimental results for these two processes~\cite{BESIII:2023sbq,BESIII:2020kpr} compared to those for $e^+e^-\to\omega\pi^0$~\cite{Achasov:2016zvn}, we opt to use $e^+e^-\to\omega\pi^0$ for the determination of $\theta_{3S-2D}$.

Table~\ref{T3} presents the results for the masses, decay widths, di-leptonic widths, and combined branching ratios of $\rho_{3S-2D}^{\prime}$, $\rho_{3S-2D}^{\prime\prime}$, $\rho_{4S-3D}^{\prime}$, and $\rho_{4S-3D}^{\prime\prime}$ calculated by taking $\theta_{3S-2D}=\pm23.4^{\circ}$ and $\theta_{3S-2D}=\pm25.1^{\circ}$, which will be used as input values for the corresponding parameters when fitting the experimental data in Sec.~\ref{Sec3B}.  Supported by theoretical studies on mass spectra and decay properties, the resonance parameters and contributions of intermediate resonant states can be determined. This will effectively limit the number of free parameters in the fitting process and avoid multiple solutions in the fitting results.

\begin{table*}[htbp]
\centering
\renewcommand\arraystretch{1.5}
\caption{Masses and decay properties of $\rho_{3S-2D}^{\prime}$, $\rho_{3S-2D}^{\prime\prime}$, $\rho_{4S-3D}^{\prime}$, and $\rho_{4S-3D}^{\prime\prime}$. Here, mixing angles for $3S$-$2D$ wave and $4S$-$3D$ wave admixtures are taken as $\pm23.4^{\circ}$ and $\pm25.1^{\circ}$, respectively. \label{T3}}
{\tabcolsep0.12in
\begin{tabular}{ccccc|cccc}
\toprule[1pt]\toprule[1pt]

 & \multicolumn{4}{c}{Positive angle } & \multicolumn{4}{c}{Negative angle} \\

 \cmidrule[1pt]{2-5}\cmidrule[1pt]{6-9}

Parameters  &  $\rho^{\prime}_{3S-2D}$   & $\rho^{\prime\prime}_{3S-2D}$  &  $\rho^{\prime}_{4S-3D}$   & $\rho^{\prime\prime}_{4S-3D}$   &  $\rho^{\prime}_{3S-2D}$   & $\rho^{\prime\prime}_{3S-2D}$   &  $\rho^{\prime}_{4S-3D}$   & $\rho^{\prime\prime}_{4S-3D}$  \\
\midrule[1pt]
Mass (MeV)                                             & 1828      & 2034      & 2150      & 2311      & 1828        & 2034                    & 2150     & 2311  \\
$\Gamma_{\rm{tot}}$ (MeV)                              & 109       & 243       & 103       & 182       & 97          & 207                     & 77       & 192  \\
$\Gamma_{e^+e^-}$ (eV)                                 & 93.35     & 85.94     & 31.03     & 51.81     & 169.59      & 1.14                    & 79.21    & 0.02 \\
$\Gamma_{e^+e^-}\mathcal{B}_{a_2(1320) \pi}$ (eV)      & 5.51      &7.25       & 1.11      & 4.75      & 27.15       & $3.67\times 10^{-4}$    & 9.76     & $5.52\times 10^{-7}$ \\
 $\Gamma_{e^+e^-}\mathcal{B}_{\omega\pi^0}$ (eV)       & 6.87      &7.17       & 0.72      & 2.20      & 30.01       & $5.96\times 10^{-7}$    & 5.49     & $2.90\times 10^{-6}$  \\
$\Gamma_{e^+e^-}\mathcal{B}_{f_1(1285)\rho}$ (eV)      &$\cdots$   &$\cdots$   & 0.63      & 0.11      & $\cdots$    &$\cdots$                 &0.92      & $3.28\times 10^{-4}$ \\
 $\Gamma_{e^+e^-}\mathcal{B}_{\pi\pi}$ (eV)            & 12.70     & 2.61      & 2.74      & 1.00      & 1.38        & 0.14                    & 0.46     & $2.01\times 10^{-3}$  \\
 $\Gamma_{e^+e^-}\mathcal{B}_{\rho\eta}$ (eV)          & 1.29      & 2.39      & 0.12      & 0.58      & 5.63        & $1.01\times 10^{-6}$    & 0.89     & $7.35\times10^{-7}$   \\
 $\Gamma_{e^+e^-}\mathcal{B}_{\rho\eta'}$ (eV)         & 0.02      & 0.13      & 0.02      & 0.02      & 0.12        & $1.76\times 10^{-6}$    & 0.03     & $2.62\times10^{-7}$    \\
\bottomrule[1pt]\bottomrule[1pt]
\end{tabular}
}
\end{table*}

\subsection{Combined analysis of six isospin vector $e^+e^-$ annihilation processes}\label{Sec3B}


Analysis in Sec.~\ref{sec2} shows that the $Y(2044)$ cannot be explained solely by the interference framework without $S$-$D$ mixing.
Therefore, in Sec~\ref{sec3A}, we conducted a systematic study of the spectrum and decay properties of higher $\rho$ mesons around 2 GeV within the $S$-$D$ mixing effect. The results show that the $S$-$D$ mixing will significantly affects the spectrum and decay properties of these $\rho$ mesons. Based on this, in this subsection, we employ two fitting schemes to fit the cross section data for $e^+e^-\to a_2(1320)\pi$~\cite{BESIII:2023sbq}, $e^+e^-\to \omega\pi^0$~\cite{BESIII:2020xmw,Achasov:2016zvn}, $e^+e^-\to f_1(1285)\pi^+\pi^-$~\cite{BaBar:2007qju,BaBar:2022ahi}, $e^+e^-\to \pi^+\pi^-$~\cite{BaBar:2019kds}, $e^+e^-\to \rho \eta$~\cite{BESIII:2023sbq}, and $e^+e^-\to \eta^{\prime} \pi^+\pi^-$~\cite{BESIII:2020kpr}, to determine the sign of the mixing angle $\theta_{4S-3D}$ and to understand the nature of the $Y(2044)$. Additionally, the significant discrepancies in the resonance parameters of $\rho$-like structures around 2 GeV observed in various electron-positron annihilation processes have long been perplexing. Another objective is to attempt to understand all observed $\rho$-like structures around 2 GeV in these six processes within a unified framework.

In the fitting process, the masses, widths, and the relevant combined branching ratios $\Gamma_{e^+e^-}\mathcal{B}_{\rm{final\,states}}$ of intermediate $\rho$ mesons take the values listed in Table~\ref{T3}, and the parameters $C_0$ and $n$ in the continuum amplitude, and the interference phase $\phi_k$ between the different amplitudes are treated as free parameters, which can be determined by fitting to the experimental cross section data. For processes $e^+e^-\to f_1(1285)\pi^+\pi^-$ and $e^+e^-\to \eta^{\prime}\pi^+\pi^-$, the two-body phase space $\Phi_{2}(s)$ in Eq. (\ref{eq1}) should be replaced with the three-body phase space $\Phi_{3}(s)$. The branching ratios $\mathcal{B}_{f_1(1285)\pi^+\pi^-}$ and $\mathcal{B}_{\eta^{\prime}\pi^+\pi^-}$ for $\rho_{3S-2D}^{\prime}$, $\rho_{3S-2D}^{\prime\prime}$, $\rho_{4S-3D}^{\prime}$ and $\rho_{4S-3D}^{\prime\prime}$ are calculated using the effective Lagrangian approach, with the results shown in Table~\ref{T4}. The amplitudes involved in the calculation can be found in Refs.~\cite{Zhou:2022ark,Liu:2022yrt}, and the coupling constants are derived from the two-body decay properties presented in Table~\ref{T4}.

\begin{table*}[htbp]
\centering
\renewcommand\arraystretch{1.5}
\caption{The the coupling constants of $g_{\rho^*f_1(1285)\rho}$ and $g_{\rho^*\eta^{\prime}\rho}$, and branching ratios of $\rho^*$ decay to $f_1(1285)\pi^+\pi^-$ and $\eta^{\prime}\pi^+\pi^-$. Here, the values outside the parentheses represent the results with the mixing angle taken as positive, while the values inside the parentheses represent the results with the mixing angle taken as negative. \label{T4}}
{\tabcolsep0.12in
\begin{tabular}{ccccc}
\toprule[1pt]\toprule[1pt]
 Parameters  & $g_{\rho^*f_1(1285)\rho}$  ($\rm{GeV}^{-2}$) & $\Gamma_{e^+e^-}\mathcal{B}_{f_1(1285)\pi^+\pi^-}$ (eV) & $g_{\rho^*\eta^{\prime}\rho}$ ($\rm{GeV}^{-1}$)  & $\Gamma_{e^+e^-}\mathcal{B}_{\eta^{\prime}\pi^+\pi^-}$ (eV)\\
 \midrule[1pt]
$\rho^{\prime}_{3S-2D}$        & 0.69~(0.50)    & 0.20 ~(0.22)      & 0.19~(0.33)      & 0.03 ~(0.19) \\
$\rho^{\prime\prime}_{3S-2D}$  & 0.037~(0.48)   & 0.02 ~(0.04)      & 0.30~(0.01)      & 0.18 ~($2.52\times10^{-6}$) \\
$\rho^{\prime}_{4S-3D}$        & 0.12~(0.079)   & 0.93 ~(1.38)      & 0.10~(0.07)      & 0.03 ~(0.04) \\
$\rho^{\prime\prime}_{4S-3D}$  & 0.033~(0.095)  & 0.20 ~(0.0006)    & 0.08~(0.02)      & 0.03 ~($4.27\times10^{-7}$) \\
\bottomrule[1pt]\bottomrule[1pt]
\end{tabular}
}
\end{table*}

\begin{figure*}[htbp]
  \centering
  \begin{tabular}{ccc}
  \includegraphics[width=195pt]{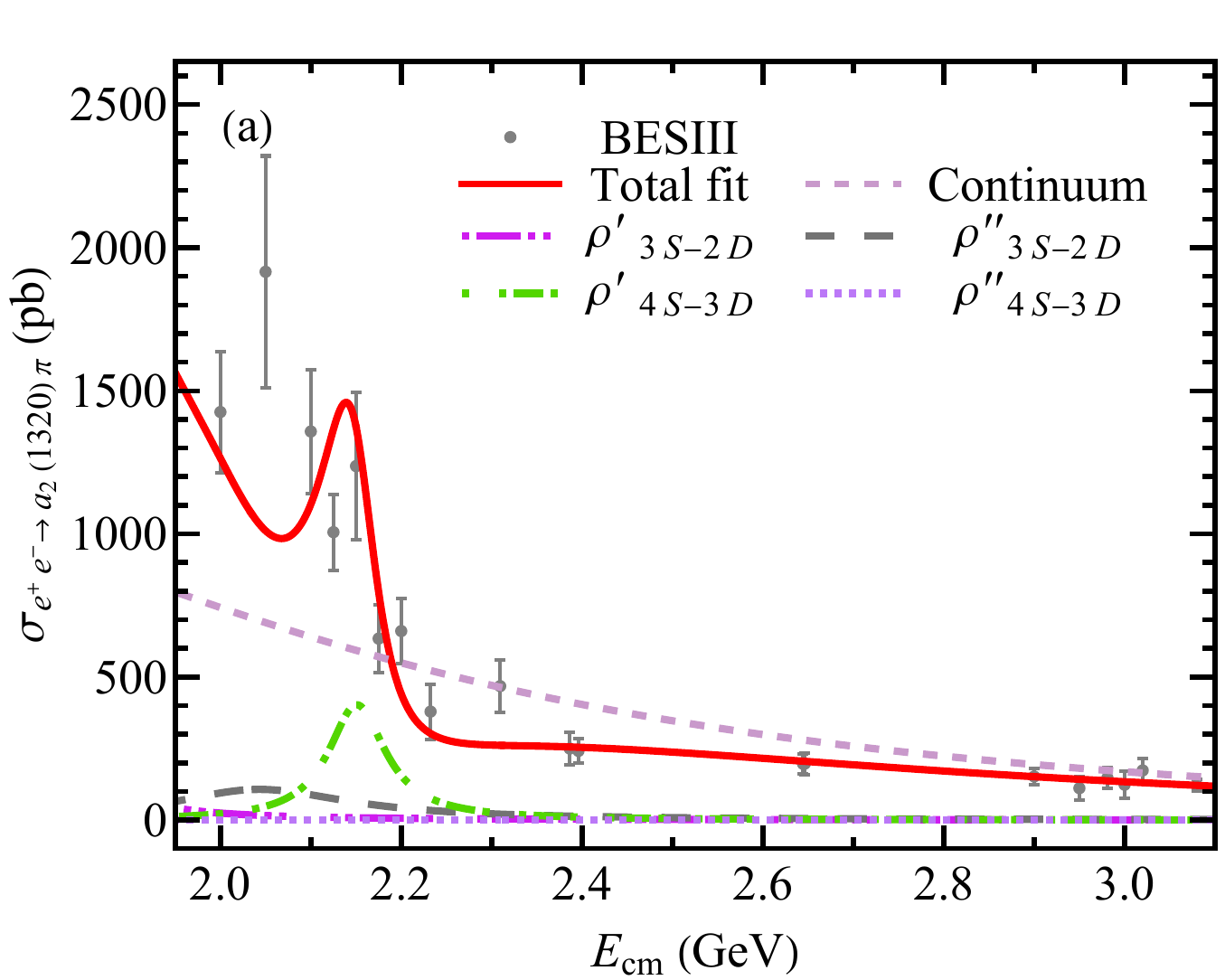}&$\quad$&\includegraphics[width=195pt]{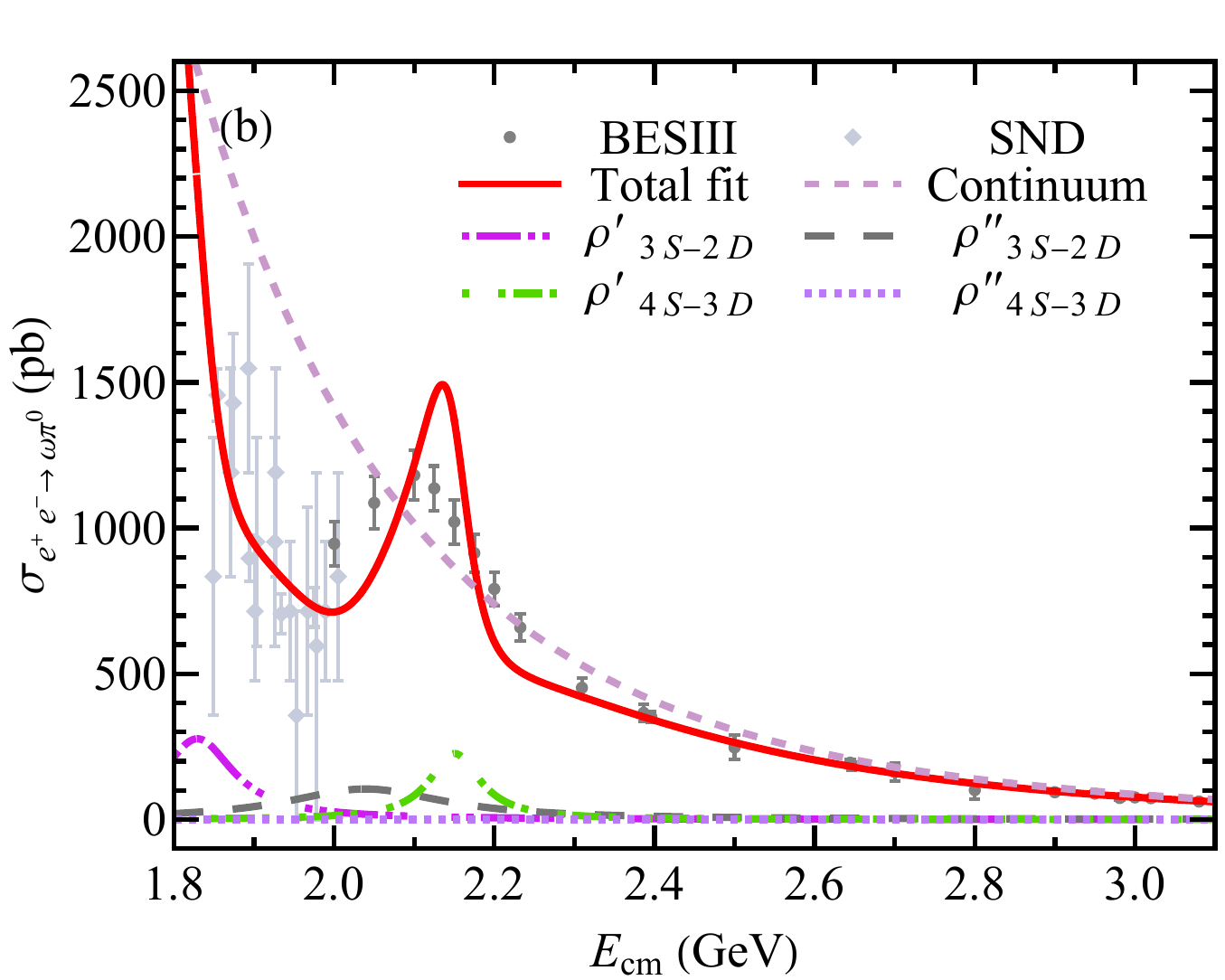}\\
  \includegraphics[width=195pt]{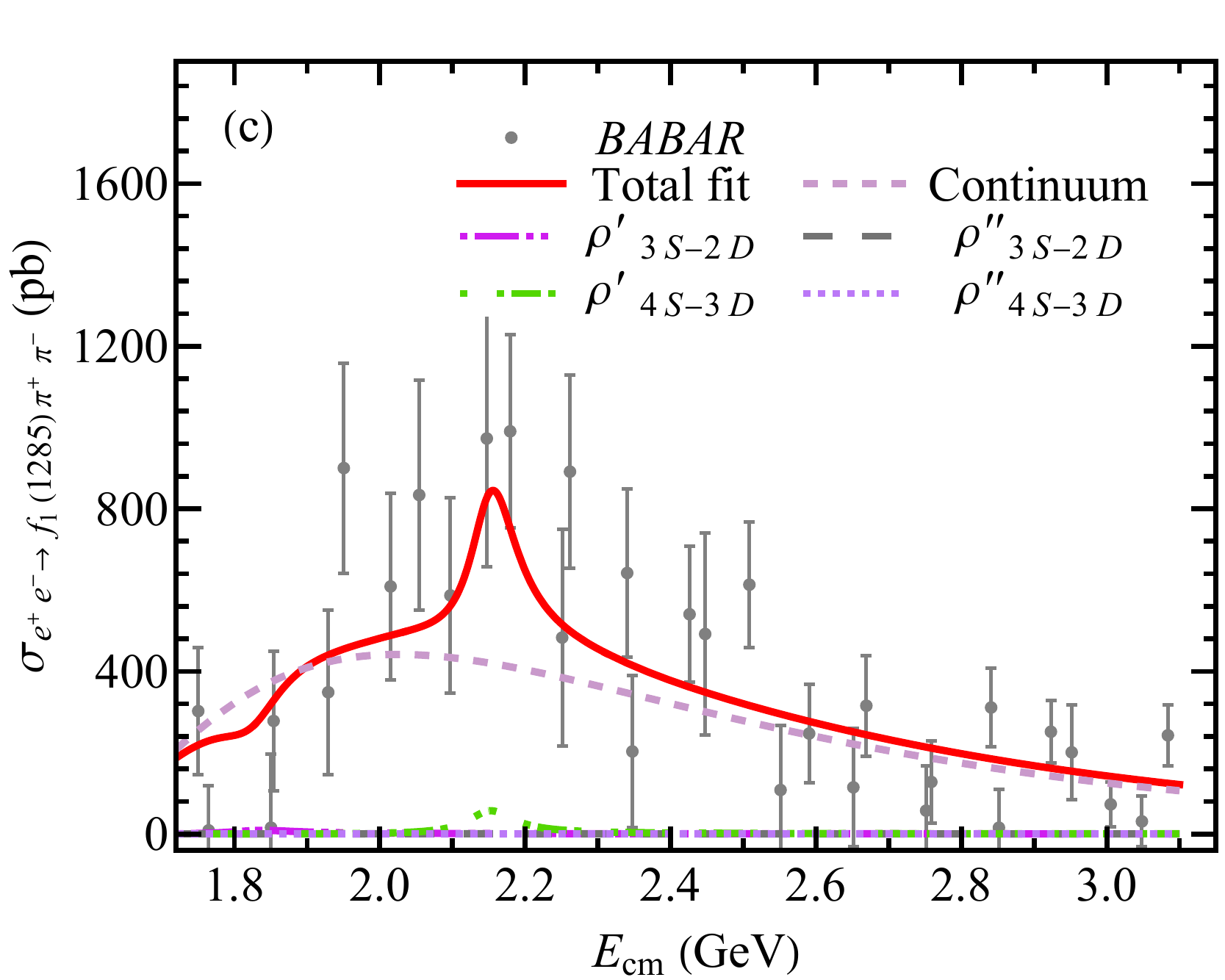}&$\quad$&\includegraphics[width=195pt]{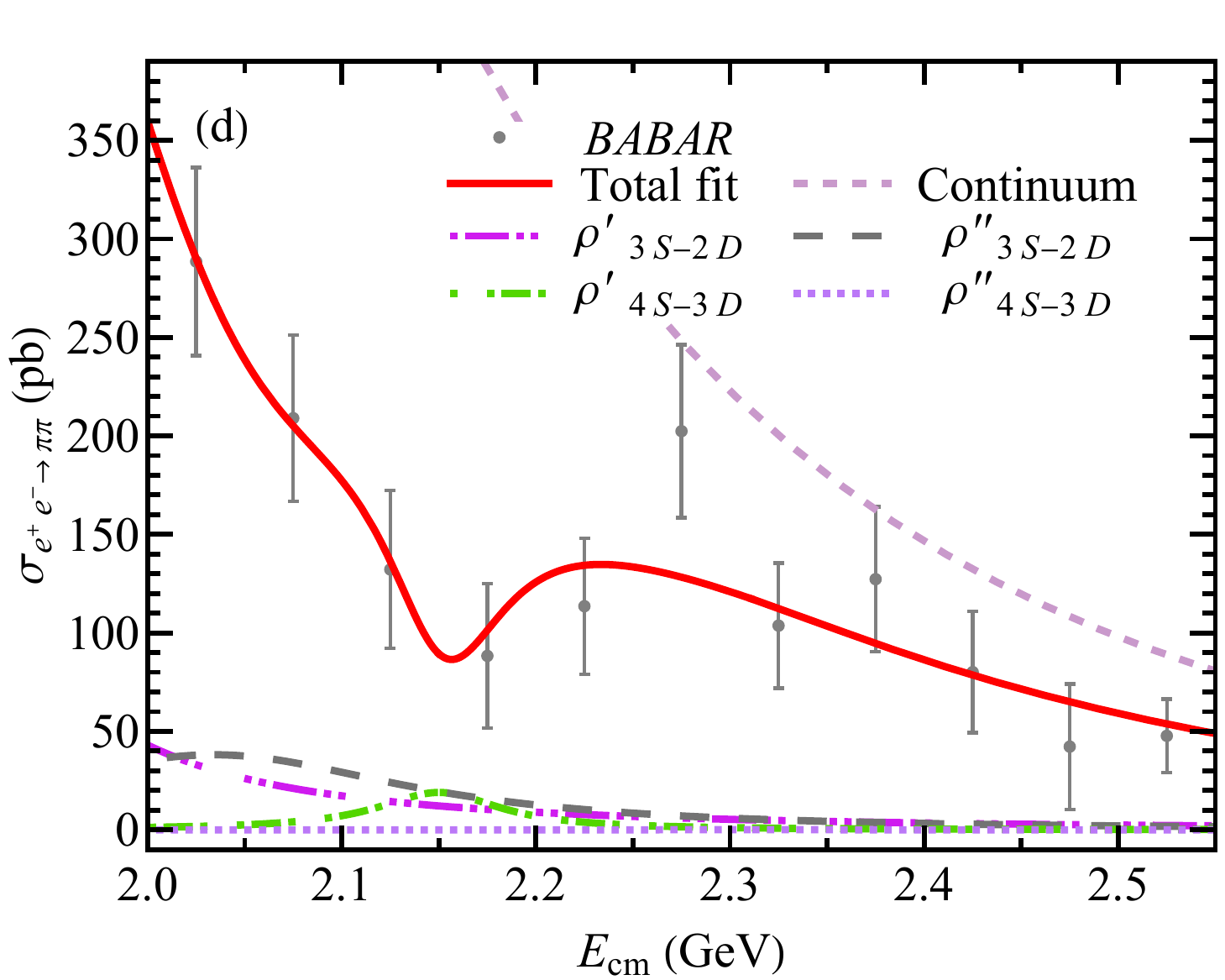}\\
  \includegraphics[width=195pt]{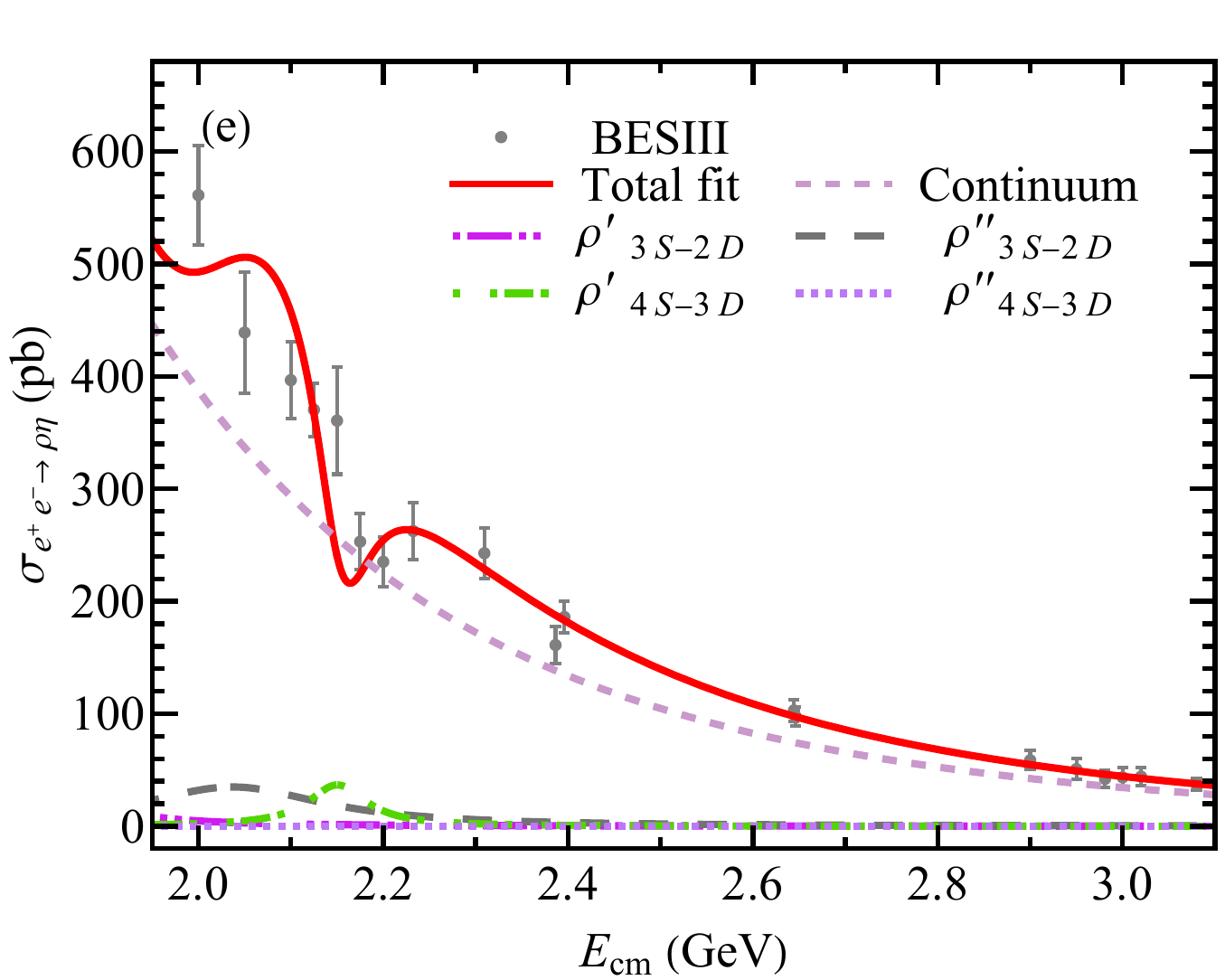}&$\quad$&\includegraphics[width=195pt]{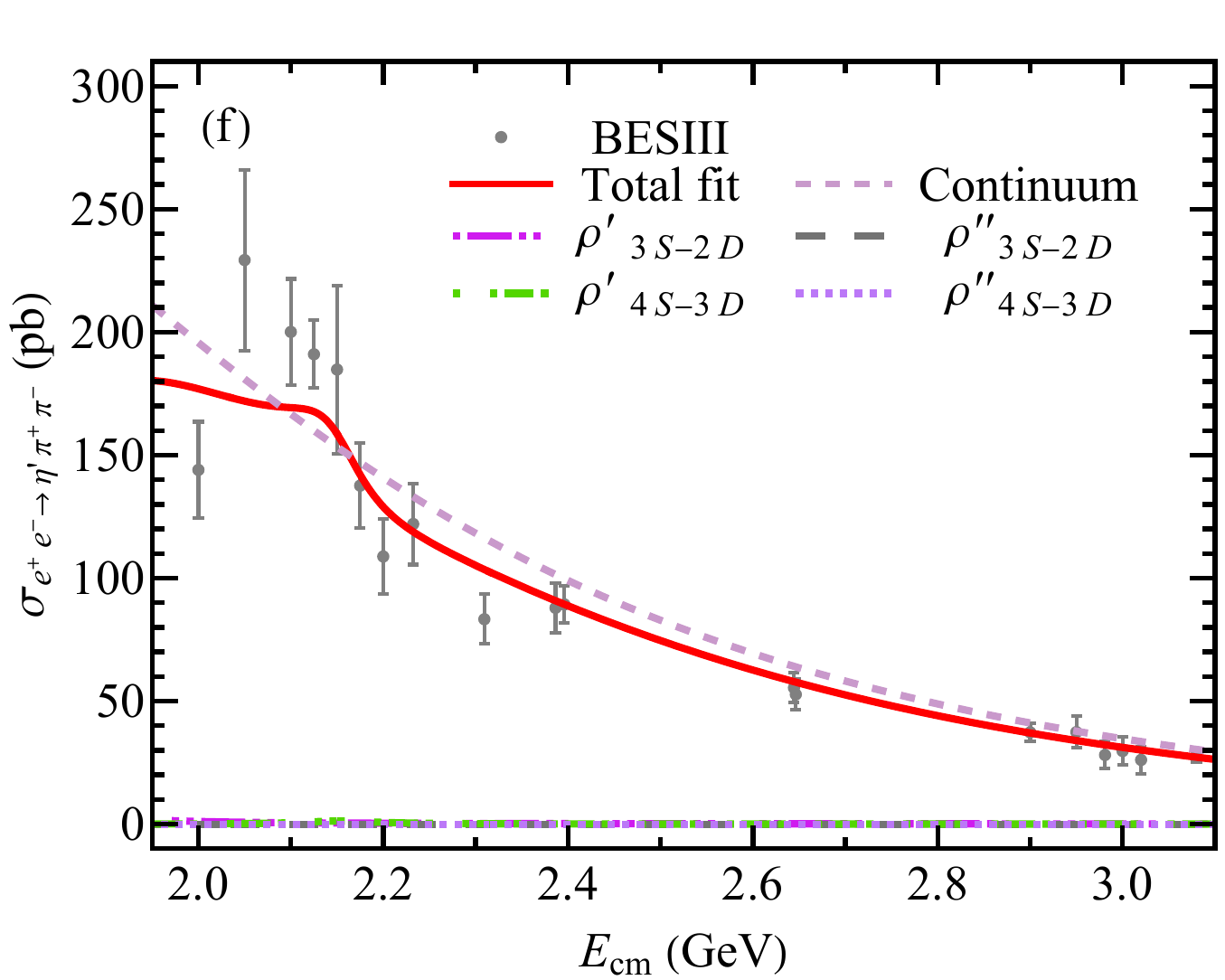}\\
  \end{tabular}
  \caption{Fit to the cross sections of the processes $e^+e^-\to a_2(1320)\pi$~\cite{BESIII:2023sbq}, $e^+e^-\to \omega\pi^0$~\cite{BESIII:2020xmw,Achasov:2016zvn}, $e^+e^-\to f_1(1285)\pi^+\pi^-$~\cite{BaBar:2007qju,BaBar:2022ahi}, $e^+e^-\to \pi^+\pi^-$~\cite{BaBar:2019kds}, $e^+e^-\to \rho \eta$~\cite{BESIII:2023sbq}, and $e^+e^-\to \eta^{\prime} \pi^+\pi^-$~\cite{BESIII:2020kpr} within the scheme 1.\label{F4}}
\end{figure*}

\begin{figure*}[htbp]
  \centering
  \begin{tabular}{ccc}
  \subfigure{\label{F5a}\includegraphics[width=195pt]{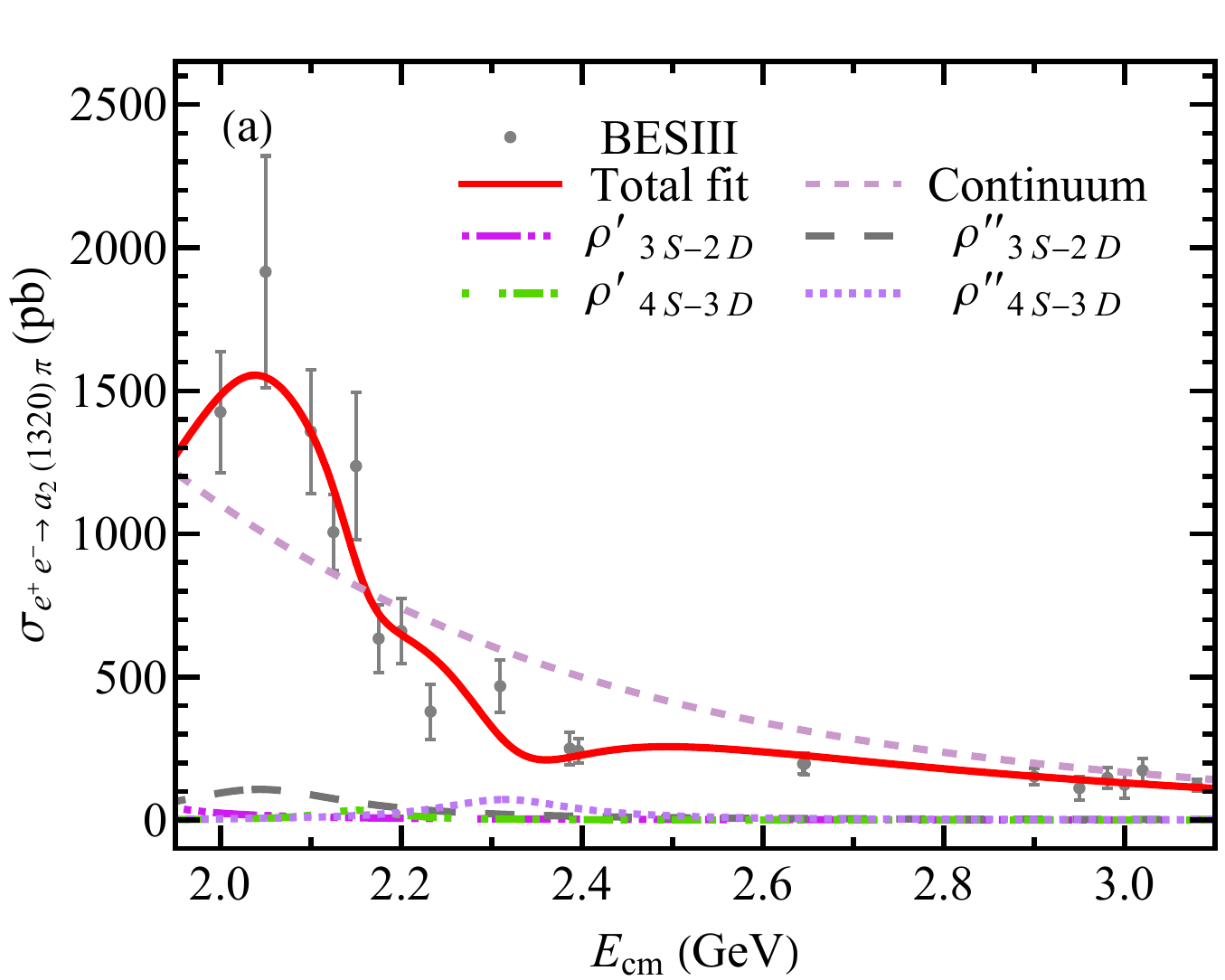}}&$\quad$&\subfigure{\label{F5b}\includegraphics[width=195pt]{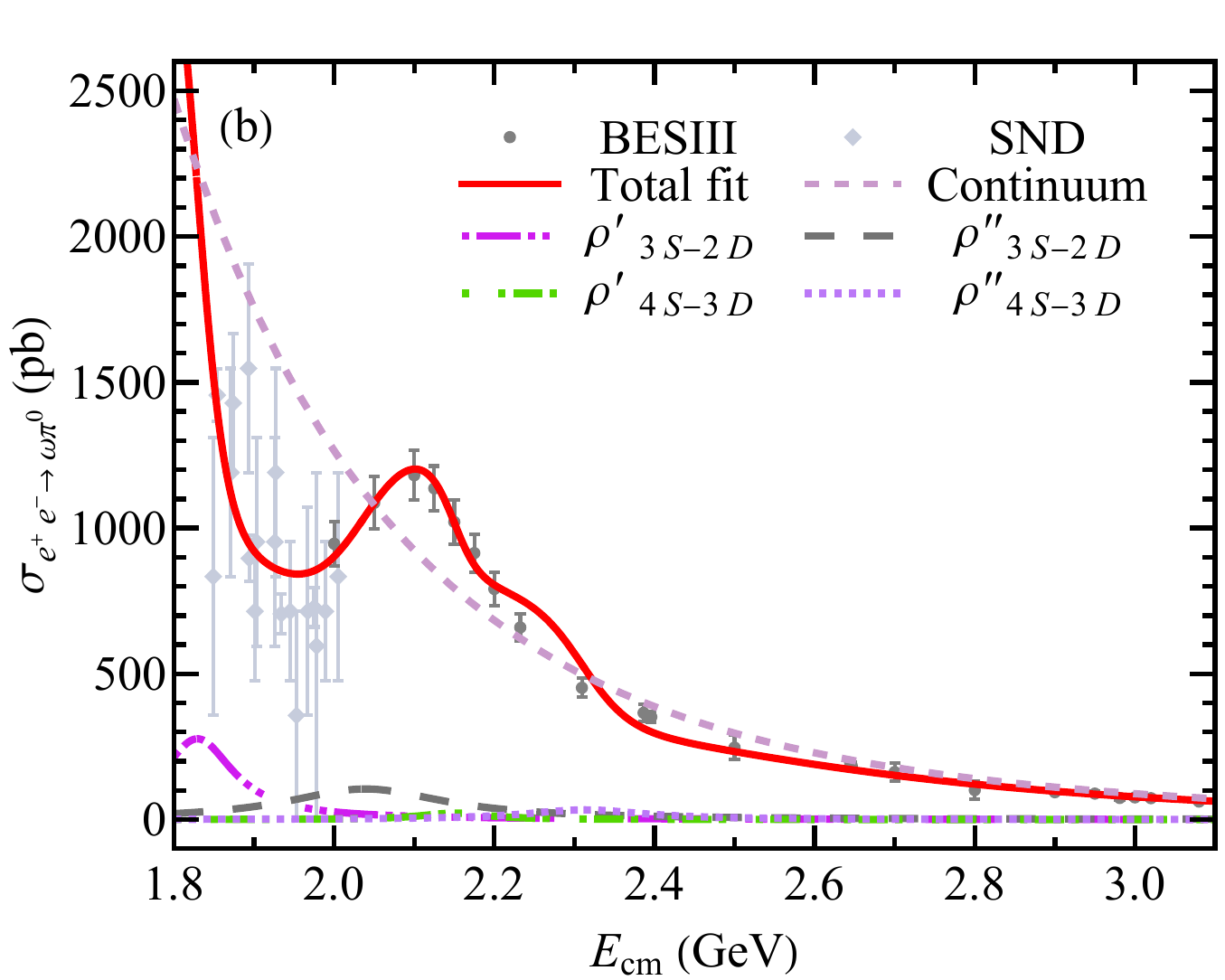}}\\
  \subfigure{\label{F5c}\includegraphics[width=195pt]{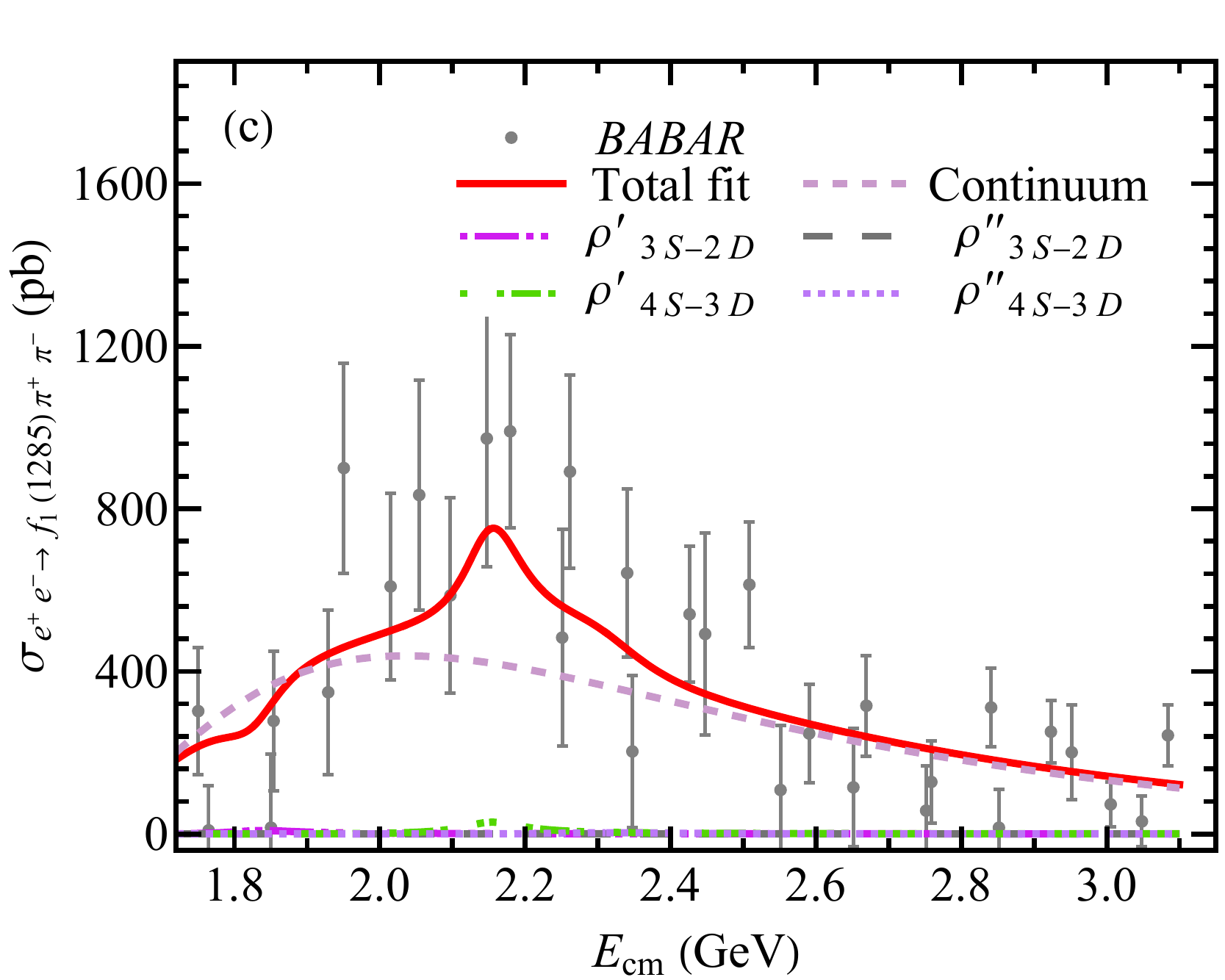}}&$\quad$&\subfigure{\label{F5d}\includegraphics[width=195pt]{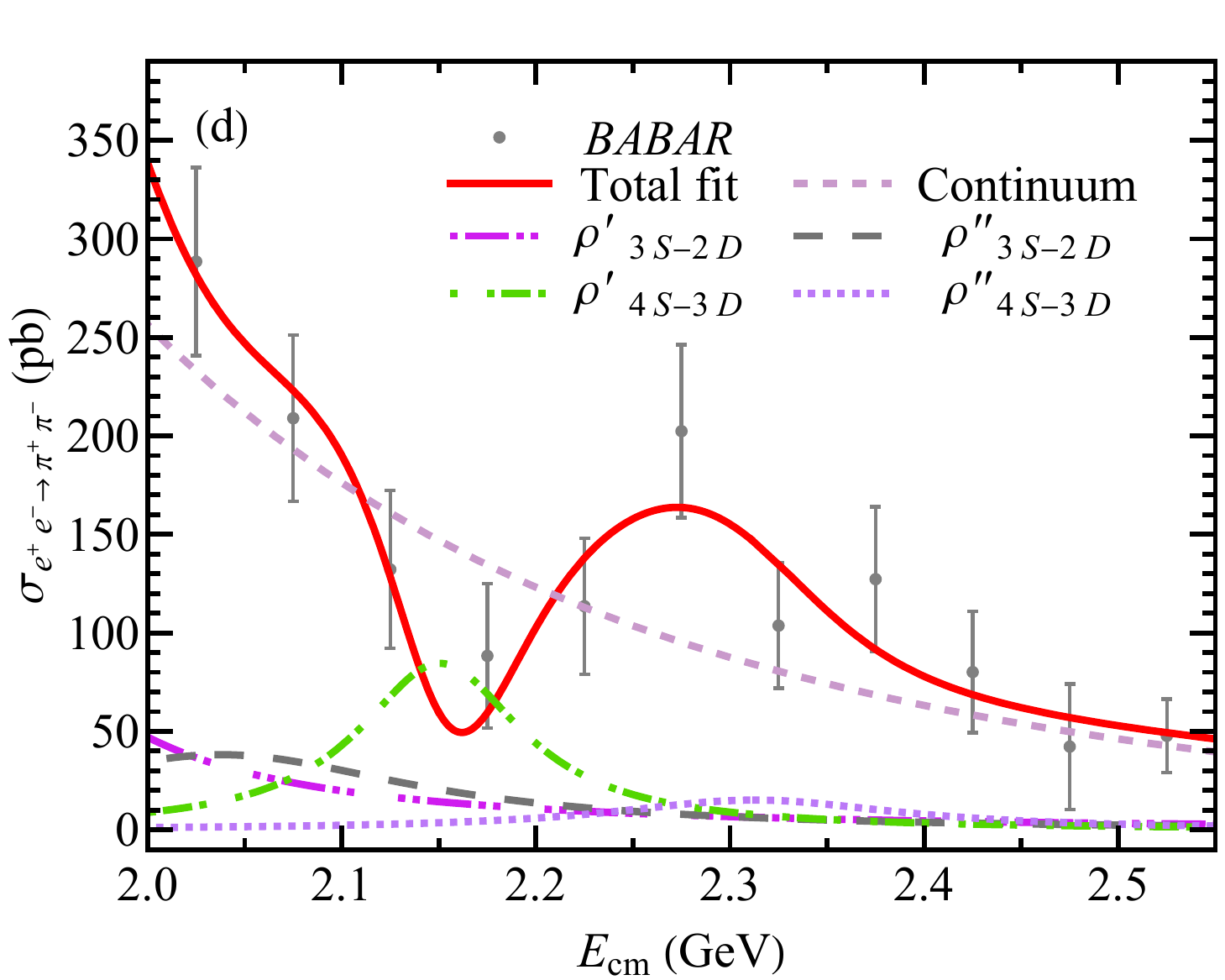}}\\
  \subfigure{\label{F5e}\includegraphics[width=195pt]{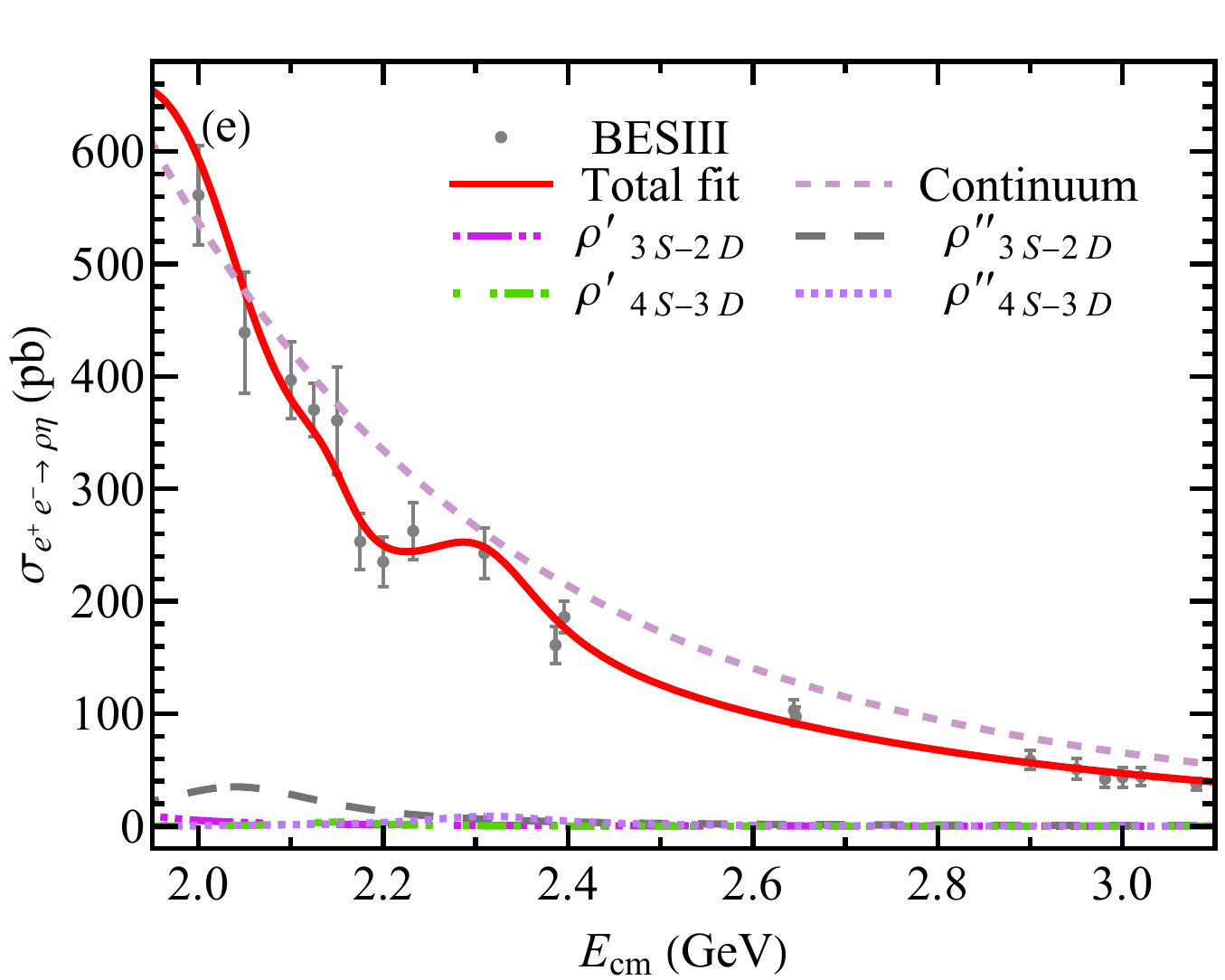}}&$\quad$&\subfigure{\label{F5f}\includegraphics[width=195pt]{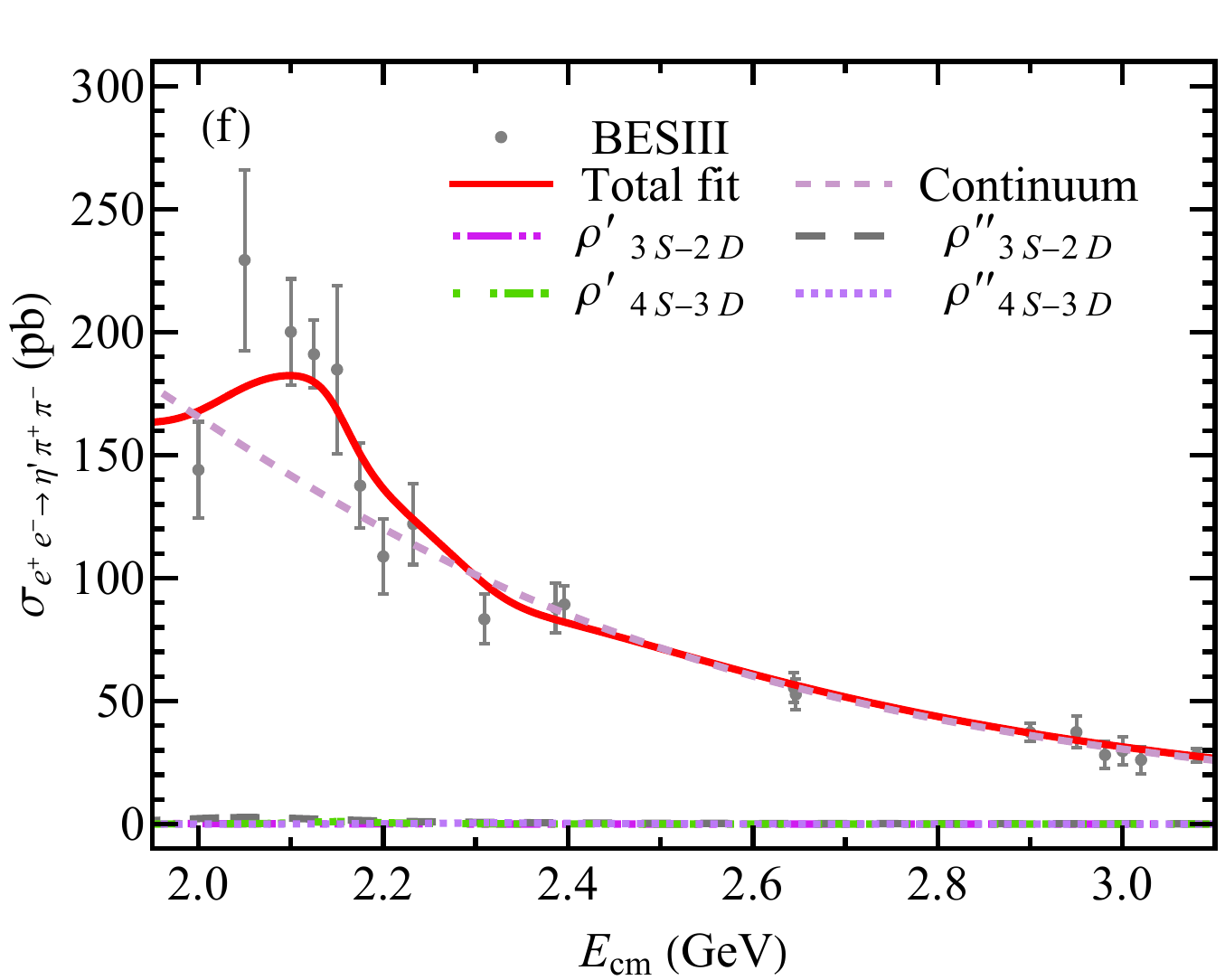}}\\
  \end{tabular}
  \caption{Fit to the cross sections of the processes $e^+e^-\to a_2(1320)\pi$~\cite{BESIII:2023sbq}, $e^+e^-\to \omega\pi^0$~\cite{BESIII:2020xmw,Achasov:2016zvn}, $e^+e^-\to f_1(1285)\pi^+\pi^-$~\cite{BaBar:2007qju,BaBar:2022ahi}, $e^+e^-\to \pi^+\pi^-$~\cite{BaBar:2019kds}, $e^+e^-\to \rho \eta$~\cite{BESIII:2023sbq}, and $e^+e^-\to \eta^{\prime} \pi^+\pi^-$~\cite{BESIII:2020kpr} within the scheme 2.\label{F5}}
\end{figure*}

The fitting results for Scheme 1 ($\theta_{4S-3D}=-25.1^{\circ}$) and Scheme 2 ($\theta_{4S-3D}=+25.1^{\circ}$) are shown in Figs.~\ref{F4} and \ref{F5}, respectively, with the relevant fitting parameters listed in Tables~\ref{T5} and \ref{T6}.
It can be seen that the scheme 1 fails to reproduce these cross section data, particularly for reproducing the cross section data of $e^+e^-\to\omega\pi^0$~\cite{BESIII:2020xmw,Achasov:2016zvn}. In contrast, the Scheme 2 can well reproduce the cross section data for all six processes. This implies that the mixing angle between $\rho(4S)$ and $\rho(3D)$ favors $\theta_{4S-3D}=+25^{\circ}$.
Therefore, in the previous subsection, it is reasonable to consider the observed $Y(2034)$ in the $e^+e^-\to \omega\pi^0$~\cite{BESIII:2020xmw} as the ideal state for $\rho_{3S-2D}^{\prime\prime}$ to determine the mixing angle $\theta_{3S-2D}$, in which $\theta_{4S-3D}>0$ is assumed.
Subsequently, we will provide a detailed discussion on the fitting results of the scheme 2.

From Fig.~\ref{F5a}, it is evident that the cross section data for the process $e^+e^-\to a_2(1320)\pi$~\cite{BESIII:2020xmw} can be well reproduced by introducing the $\rho_{3S-2D}^{\prime}$, $\rho_{3S-2D}^{\prime\prime}$, $\rho_{4S-3D}^{\prime}$, and $\rho_{4S-3D}^{\prime\prime}$ as intermediate resonance states. Although the combined branching ratios $\Gamma_{e^+e^-}\mathcal{B}_{a_2(1320) \pi}$ of the four intermediate $\rho$ meson states presented in Table~\ref{T3} are of similar magnitude, the fitting result shows that the enhancement structure near 2044 MeV is primarily originates from the the $\rho_{3S-2D}^{\prime\prime}$. This explains why the resonance parameters of $Y(2044)$ measured by the BESIII Collaboration are close to those of the $\rho_{3S-2D}^{\prime\prime}$ calculated by this work. Hence, the nature of $Y(2044)$ encompasses contributions from all four $\rho$ meson states, with the predominant contribution coming from $\rho_{3S-2D}^{\prime\prime}$.
The combined branching ratio $\Gamma_{e^+e^-}\mathcal{B}_{a_2(1320) \pi}$ of the mixed state $\rho_{3S-2D}^{\prime\prime}$ is significantly enhanced compared to that of the pure $\rho(2D)$ state due to the contribution from the component of $\rho(3S)$. Furthermore, by sufficiently considering the interference effect between the intermediate $\rho$ meson states and continuum contribution in our analysis, the previously mentioned puzzle has been effectively resolved.

The fitting result for the cross section data of process $e^+e^-\to \omega\pi^0$~\cite{BESIII:2020xmw,Achasov:2016zvn} shown in Fig.~\ref{F5b} also indicates that the resonant structure $Y(2034)$ is primarily attributed to the contribution of $\rho^{\prime\prime}_{3S-2D}$.
From the Table~\ref{T3}, we can see that the relative contribution of $\rho^{\prime\prime}_{3S-2D}$ in process $e^+e^-\to \omega\pi^0$ is greater than that in process $e^+e^-\to a_2(1320)\pi$, and the experimental measurements for process $e^+e^-\to \omega\pi^0$ exhibit higher precision. 
The width of $\rho^{\prime\prime}_{3S-2D}$ calculated based on $Y(2034)\equiv\rho^{\prime\prime}_{3S-2D}$ is 243 MeV, which is consistent with $234\pm30$ of the experimentally measured width of $Y(2034)$~\cite{BESIII:2020xmw}. This further reinforces $Y(2034)$ as an ideal candidate state for $\rho^{\prime\prime}_{3S-2D}$.
This fitting scheme not only effectively reproduces the experimental data above 2 GeV measured by the BESIII collaboration~\cite{BESIII:2020xmw}, but also accurately describe the experimental data measured by SND~\cite{Achasov:2016zvn}. The dip near 1.9 GeV appears to be a result of interference between $\rho^{\prime}_{3S-2D}$ and $\rho^{\prime\prime}_{3S-2D}$.

The fitting result of $e^+e^-\to f_1(1285)\pi^+\pi^-$~\cite{BaBar:2007qju,BaBar:2022ahi} presented in Fig.~\ref{F5c} shows that our fitting scheme basically can describe the cross section distribution of this process.
As shown in Table~\ref{T4} and Fig.~\ref{F5c}, the $\rho_{4S-3D}^{\prime}$ plays a crucial role in the $e^+e^-\to f_1(1285)\pi^+\pi^-$, while the contributions from $\rho^{\prime}_{3S-2D}$ and $\rho^{\prime\prime}_{3S-2D}$ are suppressed, making it a golden process for studying $\rho_{4S-3D}^{\prime}$. However, the current experimental data still exhibit a significant of uncertainty. It is hoped that future BESIII and BelleII experiments will provide more precise measurements for this process.

Fig.~\ref{F5d} show that the $\rho^{\prime}_{4S-3D}$ appears as a dip around 2.15 GeV in the cross section distribution of $e^+e^-\to\pi^+\pi^-$~\cite{BaBar:2019kds}, and the enhancement structure near 2.3 GeV is mainly attributed to contributions from $\rho^{\prime\prime}_{4S-3D}$. Based on the relative values of $\Gamma_{e^+e^-}\mathcal{B}_{\pi^+\pi^-}$ summarized in Table~\ref{T3},  $e^+e^-\to\pi^+\pi^-$ can be considered an ideal process for studying $\rho^{\prime}_{3S-2D}$. As a typical decay channel for the $\rho$ meson states, it is hoped that the BESIII and BelleII experiments will pay more attention to the $e^+e^- \to \pi^+\pi^-$ process, as it is vital for establishing the spectrum of the $\rho$ meson family.

The analyses from the BESIII Collaboration claim that an evidence of a dip structure around 2180 MeV was found in the cross section line shape for
$e^+e^-\to \rho \eta$ with a statistical significance of 3.0 $\sigma$~\cite{BESIII:2023sbq}. However, our fitting result for the cross section of $e^+e^-\to \rho \eta$, presented in Fig.~\ref{F5e}, shows that there is no apparent dip structure near 2180 MeV. A more detailed energy scan is required in this region to draw any conclusive statements.

The fitting result for process $e^+e^-\to \eta^{\prime}\pi^+\pi^-$~\cite{BESIII:2020kpr} shown in Fig.~\ref{F5f} indicates that the first three data points are not well reproduced in our fitting scheme. This implies that the branching ratio for $\Gamma_{e^+e^-}\mathcal{B}_{\rho\eta^{\prime}}$ calculated in this work is underestimated compared to experimental measurement. A similar puzzle has also emerged in the isospin scalar process $e^+e^-\to\omega \eta^{\prime}$~\cite{BESIII:2024qjv}, where the measured $\Gamma_{e^+e^-}\mathcal{B}_{\omega\eta^{\prime}}$ by BESIII is also greater than theoretical estimations. This puzzle necessitates future experimental and theoretical investigations.

Overall, by introducing four theoretically predicted $S$-$D$ mixing $\rho$ meson states $\rho_{3S-2D}^{\prime}$, $\rho_{3S-2D}^{\prime\prime}$, $\rho_{4S-3D}^{\prime}$, and $\rho_{4S-3D}^{\prime\prime}$ as intermediate resonance states, we have successfully reproduced the cross section data for $e^+e^-\to a_2(1320)\pi$~\cite{BESIII:2023sbq}, $e^+e^-\to \omega\pi^0$~\cite{BESIII:2020xmw,Achasov:2016zvn}, $e^+e^-\to f_1(1285)\pi^+\pi^-$~\cite{BaBar:2007qju,BaBar:2022ahi}, $e^+e^-\to \pi^+\pi^-$~\cite{BaBar:2019kds}, $e^+e^-\to \rho \eta$~\cite{BESIII:2023sbq}, and $e^+e^-\to \eta^{\prime} \pi^+\pi^-$~\cite{BESIII:2020kpr} within a unified theoretical framework. Our analysis shows that the understanding of the experimentally observed $\rho$-like structures around 2 GeV necessitates a comprehensive consideration of both the $S$-$D$ mixing and interference effects.

\section{Summary}

Recently, the BESIII Collaboration observed a $\rho$-like structure in $e^+e^-\to a_2(1320)\pi$, referred to as $Y(2044)$. The resonance parameters of $Y(2044)$ consist with theoretical predictions for the $\rho(2D)$ state, prompting BESIII to propose $Y(2044)$ as a candidate for the $n^{2S+1}L_J = 2^3D_1$ state within the $\rho$ meson family. However, the combined branching ratio $\Gamma_{e^+e^-}\mathcal{B}_{a_2(1320)\pi}$ for $Y(2044)$ was measured to be $(34.6 \pm 17.1 \pm 6.0)$ eV or $(137.1 \pm 73.3 \pm 2.1)$ eV, which is approximately two orders of magnitude larger than $0.714$ eV of the theoretical prediction for the $\rho(2D)$~\cite{Wang:2021gle}. Thus, resolving this discrepancy is essential for understanding the nature of $Y(2044)$. 

Considering that the small mass gaps among the $\rho$ mesons near 2 GeV,  which may lead to significant interference effects in production processes, we reanalyze the cross section data of the $e^+e^-\to a_2(1320)\pi$ decay processes in Sec.~\ref{sec2} by introducing the four pure $\rho$ meson states listed in Table~\ref{T1} as intermediate states. However, our analysis shows that the $Y(2044)$ cannot be explained solely within the interference framework, which implies the possible involvement of additional dynamics. Experience from previous studies of charmonium, bottomonium, and $\omega$ meson states~\cite{Wang:2019mhs,Wang:2022jxj,Wang:2023zxj,Peng:2024xui,Li:2021jjt,Bai:2022cfz,Li:2022leg,Liu:2024ets,Liu:2023gtx,Bai:2025knk}. indicate that the $S$-$D$ mixing significantly affects the mass spectrum and decay properties of higher $\rho$ meson states. Therefore, in Sec.~\ref{sec3}, we systematically investigate the spectrum and decay properties of $\rho$ mesons around 2 GeV within the $S$-$D$ mixing framework, and select ideal channels to determine the mixing angles $\theta_{3S-2D}=23.4^{\circ}$ and $\theta_{4S-3D}=25.1^{\circ}$. We then reanalyzed six isospin vector processes $e^+e^-\to a_2(1320)\pi$~\cite{BESIII:2023sbq}, $e^+e^-\to \omega\pi^0$~\cite{BESIII:2020xmw,Achasov:2016zvn}, $e^+e^-\to f_1(1285)\pi^+\pi^-$~\cite{BaBar:2007qju,BaBar:2022ahi}, $e^+e^-\to \pi^+\pi^-$~\cite{BaBar:2019kds}, $e^+e^-\to \rho \eta$~\cite{BESIII:2023sbq}, and $e^+e^-\to \eta^{\prime} \pi^+\pi^-$~\cite{BESIII:2020kpr}, and these cross section data were well reproduced by introducing four theoretically predicted $\rho$ mesonic states $\rho_{3S-2D}^{\prime}$, $\rho_{3S-2D}^{\prime\prime}$, $\rho_{4S-3D}^{\prime}$, and $\rho_{4S-3D}^{\prime\prime}$ as intermediate resonances. Our results show the nature of $Y(2044)$ encompasses contributions from all four $\rho$ meson states, with the predominant contribution coming from $\rho_{3S-2D}^{\prime\prime}$. Additionally, we found that the $Y(2034)$ observed in process $e^+e^-\to \omega\pi^0$~\cite{BESIII:2020xmw} can be regarded as a good candidate for $\rho_{3S-2D}^{\prime\prime}$, and the
$\rho(2150)$ observed in the process $e^+e^-\to f_1(1285)\pi^+\pi^-$~\cite{BaBar:2007qju} can serve as a candidate for $\rho_{4S-3D}^{\prime}$. Furthermore, we have identified several golden channels for investigating these four $\rho$ mesons around 2 GeV, and we hope experiments like BESIII and Belle II to focus more on these channels.

The properties of light vector meson states around 2 GeV have long been poorly understood, impeding progress in exploring higher vector meson spectra. Our work not only provide an explanation for the nature of $Y(2044)$, but also offer a unified framework to account for all observed $\rho$-like structures near 2 GeV in the $e^+e^-$ annihilation processes, which advances the systematic understanding of the $\rho$ meson family and offers critical insights for disentangling its higher-mass excitations.
Our analysis suggests that the $S$-$D$ mixing and interference effect may be crucial for understanding the mass spectrum and decay behaviors of the highly excited $\rho$ meson states. This is also highly instructive for understanding the $\omega$ and $\phi$ meson states around 2 GeV and higher energy regions.

\section*{Acknowledgments}

This work is supported by the Natural Science Foundation of Inner Mongolia Autonomous Region (Grant No.2025QN01045), the Research Support Program for High-Level Talents at the Autonomous Region level in the Inner Mongolia Autonomous Region (Grant No. 13100-15112049), and the Research Startup Project of Inner Mongolia University (Grant No. 10000-23112101/101). X.L. is supported by the National Natural Science Foundation of China under Grant Nos. 12335001, and 12247101, the 111 Center under Grant No. B20063, the Natural Science Foundation of Gansu Province (No. 22JR5RA389, No. 25JRRA799), the Talent Scientific Fund of Lanzhou University, the fundamental Research Funds for the Central Universities (No. lzujbky-2023-stlt01), and the project for top-notch innovative talents of Gansu province. J. Z.W. is also supported by the Start-up Funds of Chongqing University.
H. X. is also supported by the National Natural Science Foundation of China under Grant Nos. 12465016 and 12005168, the Natural Science Foundation of Gansu Province (No. 22JR5RA171).

\begin{table*}[htbp]
\centering
\renewcommand\arraystretch{1.5}
\caption{Parameters obtained from the fit to the cross section distributions of the processes $e^+e^-\to a_2(1320)\pi$~\cite{BESIII:2023sbq}, $e^+e^-\to \omega\pi^0$~\cite{BESIII:2020xmw,Achasov:2016zvn}, $e^+e^-\to f_1(1285)\pi^+\pi^-$~\cite{BaBar:2007qju,BaBar:2022ahi}, $e^+e^-\to \pi^+\pi^-$~\cite{BaBar:2019kds}, $e^+e^-\to \rho \eta$~\cite{BESIII:2023sbq}, and $e^+e^-\to \eta^{\prime} \pi^+\pi^-$~\cite{BESIII:2020kpr} within the scheme 1.\label{T5}}
{\tabcolsep0.1in
\begin{tabular}{cccccccc}
\toprule[1pt]\toprule[1pt]
Reactions   &  $n$  &   $C_0$      &  $\phi_1$    & $\phi_2$   &  $\phi_3$    & $\phi_4$   & $\chi^2/\rm{n.d.f}$ \\
\midrule[1pt]

$e^+e^-\to a_2(1320) \pi$          & $1.39\pm0.01$       & $404.65\pm8.41$       & $0.58\pm0.72$       & $3.62\pm0.14$      & $2.51\pm0.12$       & $2.06\pm1.78$     & 2.15 \\

 $e^+e^-\to \omega\pi^0$           & $2.04\pm 0.01$       & $1105.60\pm7.07$        & $3.36\pm0.06$       & $5.32\pm0.04$       & $2.24\pm0.06$       & $1.47\pm0.65$     & 2.79  \\

$e^+e^-\to f_1(1285)\pi^+\pi^-$    & $2.29\pm 0.02$       & $6590.15\pm279.53$        & $5.30\pm0.52$       & $0.93\pm1.27$       & $1.21\pm0.46$       & $2.21\pm1.93$     & 1.63  \\

 $e^+e^-\to \pi\pi$              & $2.47\pm 0.02$       & $2065.86\pm48.67$        & $3.33\pm0.52$       & $4.42\pm0.22$       & $4.31\pm0.31$       & $1.36\pm3.14$     & 0.97  \\

 $e^+e^-\to \rho\eta$            & $1.61\pm 0.01$       & $478.52\pm5.78$        & $0.32\pm0.63$       & $1.93\pm0.09$       & $4.18\pm0.10$       & $3.66\pm5.09$     & 1.50  \\

 $e^+e^-\to \eta'\pi^+\pi^-$     & $2.09\pm 0.01$       & $1775.41\pm13.96$        & $3.77\pm0.24$       & $0.39\pm3.63$       & $2.37\pm0.21$       & $3.60\pm0.88$     & 1.46  \\
\bottomrule[1pt]\bottomrule[1pt]
\end{tabular}
}
\end{table*}

\newpage
\begin{table*}[htbp]
\centering
\renewcommand\arraystretch{1.2}
\caption{Parameters obtained from the fit to the cross section distributions of the processes $e^+e^-\to a_2(1320)\pi$~\cite{BESIII:2023sbq}, $e^+e^-\to \omega\pi^0$~\cite{BESIII:2020xmw,Achasov:2016zvn}, $e^+e^-\to f_1(1285)\pi^+\pi^-$~\cite{BaBar:2007qju,BaBar:2022ahi}, $e^+e^-\to \pi^+\pi^-$~\cite{BaBar:2019kds}, $e^+e^-\to \rho \eta$~\cite{BESIII:2023sbq}, and $e^+e^-\to \eta^{\prime} \pi^+\pi^-$~\cite{BESIII:2020kpr} within the scheme 2.\label{T6}}
{\tabcolsep0.1in
\begin{tabular}{cccccccc}
\toprule[1pt]\toprule[1pt]

Reactions   &  $n$  &   $C_0$      &  $\phi_1$    & $\phi_2$   &  $\phi_3$    & $\phi_4$   & $\chi^2/\rm{n.d.f}$ \\
\midrule[1pt]

$e^+e^-\to a_2(1320) \pi$          & $1.64\pm0.01$       & $697.68\pm8.60$       & $3.55\pm0.72$       & $1.48\pm0.14$      & $3.99\pm0.14$       & $4.20\pm0.14$     & 1.24 \\

 $e^+e^-\to \omega\pi^0$           & $1.95\pm 0.01$       & $928.09\pm3.33$        & $3.14\pm0.06$       & $6.11\pm0.02$       & $3.17\pm0.04$       & $3.13\pm0.06$     & 1.45  \\

$e^+e^-\to f_1(1285)\pi^+\pi^-$    & $2.25\pm 0.02$       & $6201.76\pm276.00$        & $5.29\pm0.68$       & $0.92\pm1.54$       & $1.31\pm0.52$       & $2.29\pm0.88$     & 1.62  \\

 $e^+e^-\to \pi^+\pi^-$            & $2.17\pm 0.03$       & $525.33\pm23.00$        & $6.25\pm0.51$       & $4.91\pm0.18$       & $4.07\pm0.14$       & $2.25\pm0.26$     & 0.86  \\

 $e^+e^-\to \rho\eta$              & $1.66\pm 0.01$       & $430.72\pm4.20$        & $3.81\pm0.39$       & $3.05\pm0.07$       & $2.67\pm0.23$       & $1.84\pm0.16$     & 0.79  \\

 $e^+e^-\to \eta'\pi^+\pi^-$       & $2.06\pm 0.01$       & $1579.80\pm13.74$        & $3.65\pm0.63$       & $0.34\pm0.10$       & $2.54\pm0.25$       & $4.10\pm0.44$     & 1.08  \\
\bottomrule[1pt]\bottomrule[1pt]
\end{tabular}
}
\end{table*}

\end{document}